\documentclass{elsart}

 \usepackage{graphicx}

\usepackage{amssymb}

\newcommand{\degree}{^{\circ}}

\begin{document}

\begin{frontmatter}



\title{Scientific performances of the XAA1.2 front-end chip for silicon microstrip detectors}


\author[IASF_Rm]{Ettore Del Monte\corauthref{Del_Monte}},
\corauth[Del_Monte]{Corresponding author. Address: Istituto di
Astrofisica Spaziale e Fisica Cosmica, INAF, Via Fosso del Cavaliere
100, I-00133 Roma, Italy. Tel: +39 06 4993 4430. Fax: +39 06 2066
0188.} \ead{ettore.delmonte@iasf-roma.inaf.it}
\author[IASF_Rm]{Paolo Soffitta},
\author[IASF_Bo]{Ennio Morelli},
\author[IASF_Rm]{Luigi Pacciani},
\author[IASF_Rm]{Geiland Porrovecchio},
\author[IASF_Rm]{Alda Rubini},
\author[IASF_Rm]{Olga Uberti},
\author[IASF_Rm]{Enrico Costa},
\author[IASF_Rm]{Giuseppe Di Persio},
\author[IASF_Rm]{Immacolata Donnarumma},
\author[IASF_Rm]{Yuri Evangelista},
\author[IASF_Rm]{Marco Feroci},
\author[IASF_Rm]{Francesco Lazzarotto},
\author[IMIP_Rm]{Marcello Mastropietro},
\author[ENEA]{Massimo Rapisarda}

\address[IASF_Rm]{Istituto di Astrofisica Spaziale e Fisica Cosmica, INAF,
Roma,\\
Via Fosso del Cavaliere 100, I-00133 Roma, Italy}
\address[IASF_Bo]{Istituto di Astrofisica Spaziale e Fisica Cosmica,
INAF, Bologna, \\
Via P. Gobetti 101, I-40129 Bologna, Italy}
\address[IMIP_Rm]{Istituto dei Sistemi Complessi, CNR, Roma,\\
Via Salaria km 29.300, I-00016 Monterotondo Scalo (RM) c.p. 10,
Italy}
\address[ENEA]{ENEA C.R. Frascati, \\
Via Enrico Fermi 45, I-00044 Frascati (RM), Italy}

\begin{abstract}

The XAA1.2 is a custom ASIC chip for silicon microstrip detectors
adapted by Ideas for the SuperAGILE instrument on board the AGILE
space mission. The chip is equipped with 128 input channels, each
one containing a charge preamplifier, shaper, peak detector and
stretcher. The most important features of the ASIC are the extended
linearity, low noise and low power consumption. The XAA1.2 underwent
extensive laboratory testing in order to study its commandability
and functionality and evaluate its scientific performances. In this
paper we describe the XAA1.2 features, report the laboratory
measurements and discuss the results emphasizing the scientific
performances in the context of the SuperAGILE front-end electronics.

\end{abstract}

\end{frontmatter}


\section{Introduction}
\label{sec:intro}

The new generation of tracking, vertex and pixel detectors for High
Energy Physics experiments, as for example the ATLAS and CMS
detectors at LHC, is based on silicon microstrip detectors. This
type of sensors offers a high spatial resolution combined with a
wide surface, a good signal-to-noise ratio and a fast response.
Standard silicon strip detectors, now common in High Energy Physics
experiments, can be successfully used for position-sensitive
detection of X-rays. A similar type of detectors is suitable in many
applications and experimental techniques in solid-state physics,
material science and biological study using synchrotron radiation,
as shown for example in \cite{Dabrowski_2003}.

Space Astrophysics is now starting to take advantage of the peculiar
features of the silicon microstrip detectors, mainly the possibility
of building large and light detectors with extremely high spatial
resolution, as proposed in \cite{Barbiellini_1987} and
\cite{Barbier_et_al_1998}. Particularly, replacing the gaseous
detectors with solid-state detectors allows to increase the lifetime
of the space missions, increase the spatial resolution up to $\mu$m
level for on-axis experiments and to tens of $\mu$m in case of
inclined penetration and reduce the deadtime from ms to $\mu$s
(\cite{do_Couto_2001}). The detectors of PAMELA
(\cite{Boezio_et_al_2004}), space experiment for Cosmic Ray Physics
launched in spring 2006, AGILE (\cite{Tavani_et_al_2006}) and GLAST
(\cite{Ritz_et_al_2004}), the next satellite-borne missions for High
Energy Astrophysics in the X-ray and gamma-ray energy bands, are all
based on the silicon microstrip technology.

The use of solid state detectors, in both Space and ground based
experiments, allows to increase the number and density of
independent channels number and density, thus requiring the ASIC
technology in the integrated front-end electronics. Critical
parameters for the front-end electronics of the silicon microstrip
detectors, in both ground and space experiments, are noise, speed,
power consumption and radiation tolerance. The noise introduced by
the detector depends on the leakage current shot noise, bias
resistors thermal noise, metal strips Johnson-type noise and
statistic fluctuation of the charge carriers. Depending on the
energy of the interacting particle, Equivalent Noise Charge values
of the order of $10^2-10^4$ $\mathrm{e^-}$ can be achieved. The
bunch crossing time in High Energy Physics and the flux variability
in Astrophysics experiments require a signal processing time of the
order of 1 $\mu$s. The amount of power available from solar panels
in a satellite-borne experiment and the difficulties associated with
supplying and cooling the electronics in an accelerator experiment
require power consumption of the order of 0.1--1 mW per input
channel. Finally, the levels of total ionizing dose, both in
accelerators and in Space, require radiation tolerance in the
krad--Mrad range. A more complete view of the requirements of the
readout of silicon microstrip detectors and the architectural and
technological options may be found in \cite{Dabrowski_2003}.

In this paper the functional and scientific performances of the
XAA1.2, as the front-end electronics circuit of the SuperAGILE
instrument, are reported. The structure of the paper is as follows:
section \ref{sec:SuperAGILE} contains the overview of the SuperAGILE
instrument and the AGILE mission, in section \ref{sec:XA_features}
the description of the XAA1.2 electronic circuit is reported,
section \ref{sec:set-up} deals with the description of the
experimental set-up, in section \ref{sec:noise_th} we discuss the
most important sources of noise in systems based on the silicon
microstrip detectors, the results of the linearity and noise
measurements with a test pulse and X-ray calibration sources are in
section \ref{sec:linearity_noise_ecal} and
\ref{sec:linearity_noise_X-rays} respectively, the threshold
measurements are in section \ref{sec:threshold}, section
\ref{sec:bias_optimization} contains the outline of the electronic
noise reduction method by means of the signal shape adjustment, the
measurements of the thermal stability of the output signals are in
section \ref{sec:thermal}, in section \ref{sec:radiation} we give an
outlook of the XAA1.2 radiation tolerance and finally section
\ref{sec:conclusions} contains the conclusions.

\section{Overview of the SuperAGILE instrument and the AGILE mission}
\label{sec:SuperAGILE}

AGILE is a satellite-borne mission dedicated to the gamma ray and
X-ray Astrophysics. The AGILE payload contains two instruments, the
Gamma Ray Imaging detector (GRID) and the SuperAGILE X-ray monitor,
both equipped with silicon microstrip detectors. The GRID
(\cite{Prest_et_al_2003}) is sensitive between 30 MeV and 50 GeV and
is composed of a Silicon Tracker, with tungsten converters and
silicon microstrip detectors, complemented by a Mini-Calorimeter
with CsI(Tl) scintillator bars (\cite{Labanti_et_al_2006}).
SuperAGILE (\cite{Soffitta_et_al_2006}), the AGILE X-ray monitor, is
a coded aperture instrument equipped with silicon microstrip
detectors and tungsten coded mask and is sensitive to hard X-rays in
the 15--45 keV energy band. SuperAGILE is located in the field of
view of the GRID and the two instruments are designed as a single
experiment. The payload is surrounded by a plastic scintillator
Anticoincidence shield (\cite{Perotti_et_al_2006}). AGILE is an ASI
(Agenzia Spaziale Italiana) satellite mission and its launch is
planned on 2007. The satellite will be positioned on an equatorial
orbit with about 560 km altitude. Further details about the AGILE
mission and scientific objectives may be found in
\cite{Tavani_et_al_2006}. An overall description of the SuperAGILE
experiment will be reported in a forthcoming paper
\cite{Feroci_et_al_inprep}).

Among the scientific objectives of the AGILE mission are the Gamma
Ray Bursts (about 15--20 events per year are expected in the hard
X-ray band and 5--10 events per year in the gamma ray band, they
will be localized on board and their coordinates will be distributed
to the scientific community by means of the ORBCOMM\footnote{Web
site: \texttt{www.orbcomm.com}} satellite constellation), Active
Galactic Nuclei (with study of transients and low-level gamma ray
emission and duty cycles, observation of the relationship and
correlation between the gamma ray variability and the radio,
optical, X-ray and ultra high-energy emission), Compact Galactic
Sources (Gamma Ray Pulsars, Unidentified EGRET sources, black hole
candidates, X-ray binary systems and Soft Gamma-Ray Repeaters
outbursts) and Diffuse Emission (with the study of cosmic ray
origin, propagation, interaction and emission processes).

SuperAGILE can produce images of the sources in the Sky using the
coded aperture imaging principle, based on a position sensitive
detector coupled to a plane with a pattern of pixels opaque and
transparent to the radiation under study and distributed following a
specific code (``coded mask''). The X-ray sources cast unique
shadows of the mask on the detector and the photons direction can be
reconstructed using proper algorithms. More details about the coded
mask instruments, their principle and their performances as imagers
for High Energy Astrophysics may be found in
\cite{Fenimore_Cannon_1978} and \cite{Zand_1992}.

The SuperAGILE detector is composed of four modules of one dimension
silicon microstrip detector tiles (manufactured by Hamamatsu) of $19
\times 19$ cm$^2$ area and 410 $\mu$m thickness, each one containing
1536 strips. Each detector module is composed of two ladders with
768 strips and $9.5 \times 19 \; \mathrm{cm^2}$ surface, composed in
turn of two square tiles of 9.5 cm side, bonded head-to-head with
consecutive strips. The detector capacitance is 30 $\mathrm{pF \cdot
strip^{-1}}$, the pitch of the microstrips is 121 $\mu$m and above
each strip an aluminum electrode (of 72 $\Omega$ resistance) is
positioned, separately connected to an input chain of the XAA1.2.
The detector is biased at about $+90$ V on the back side. The
detector modules are glued on a honeycomb and carbon fiber tray.

\begin{figure}[ph!] \centering

\includegraphics[width=12 cm]{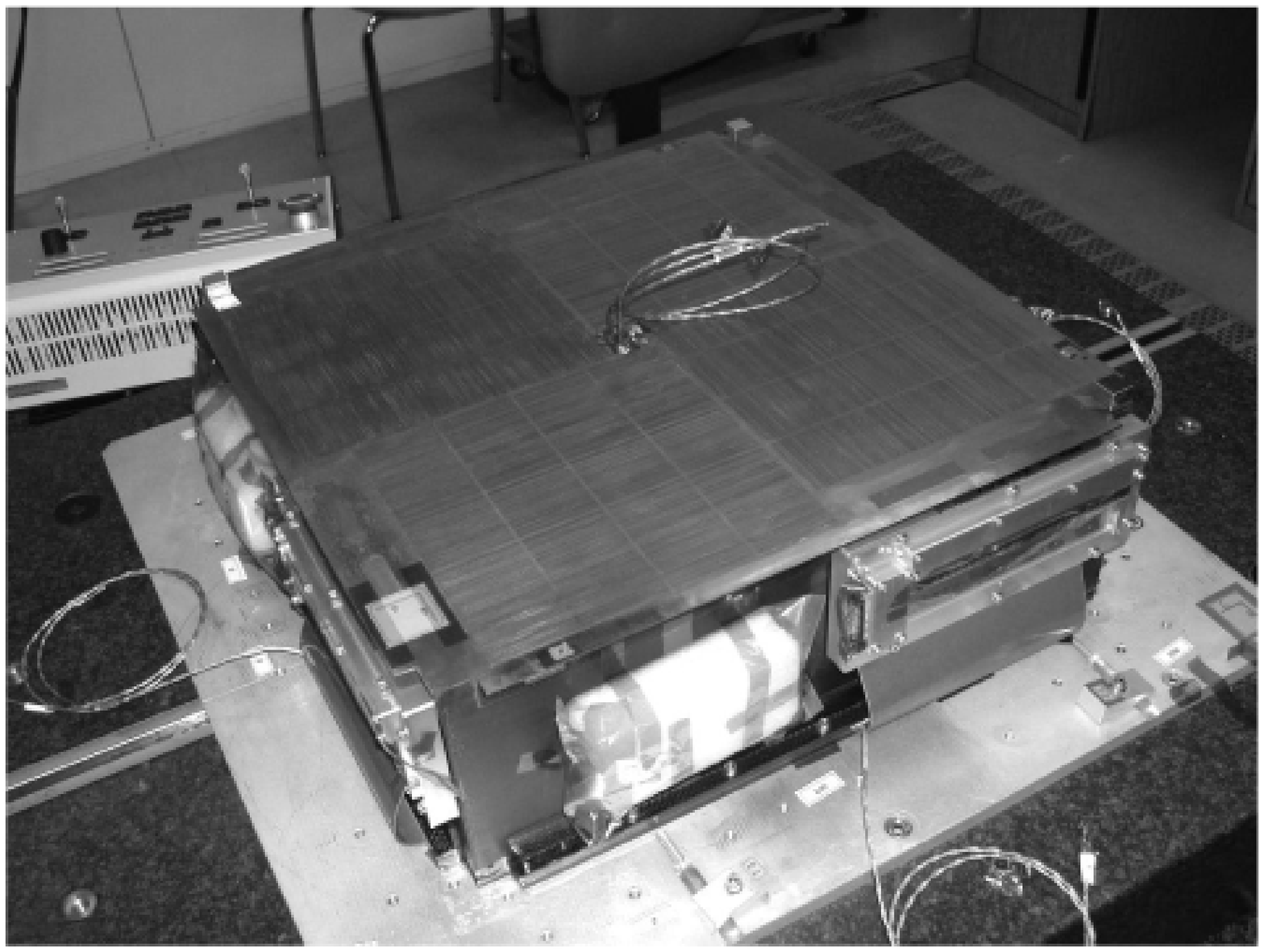}
\caption{Picture of the SuperAGILE instrument: the upper layer is
the coded mask, supported by the collimator. The silicon
microstrip detector is not visible in the picture. The vertical
SAFEE boxes are fixed to the collimator external walls. The wires
shown in the picture are used to connect the thermometers to the
dedicated conditioning electronics.} \label{fig:SuperAGILE_photo}

\includegraphics[width=12 cm]{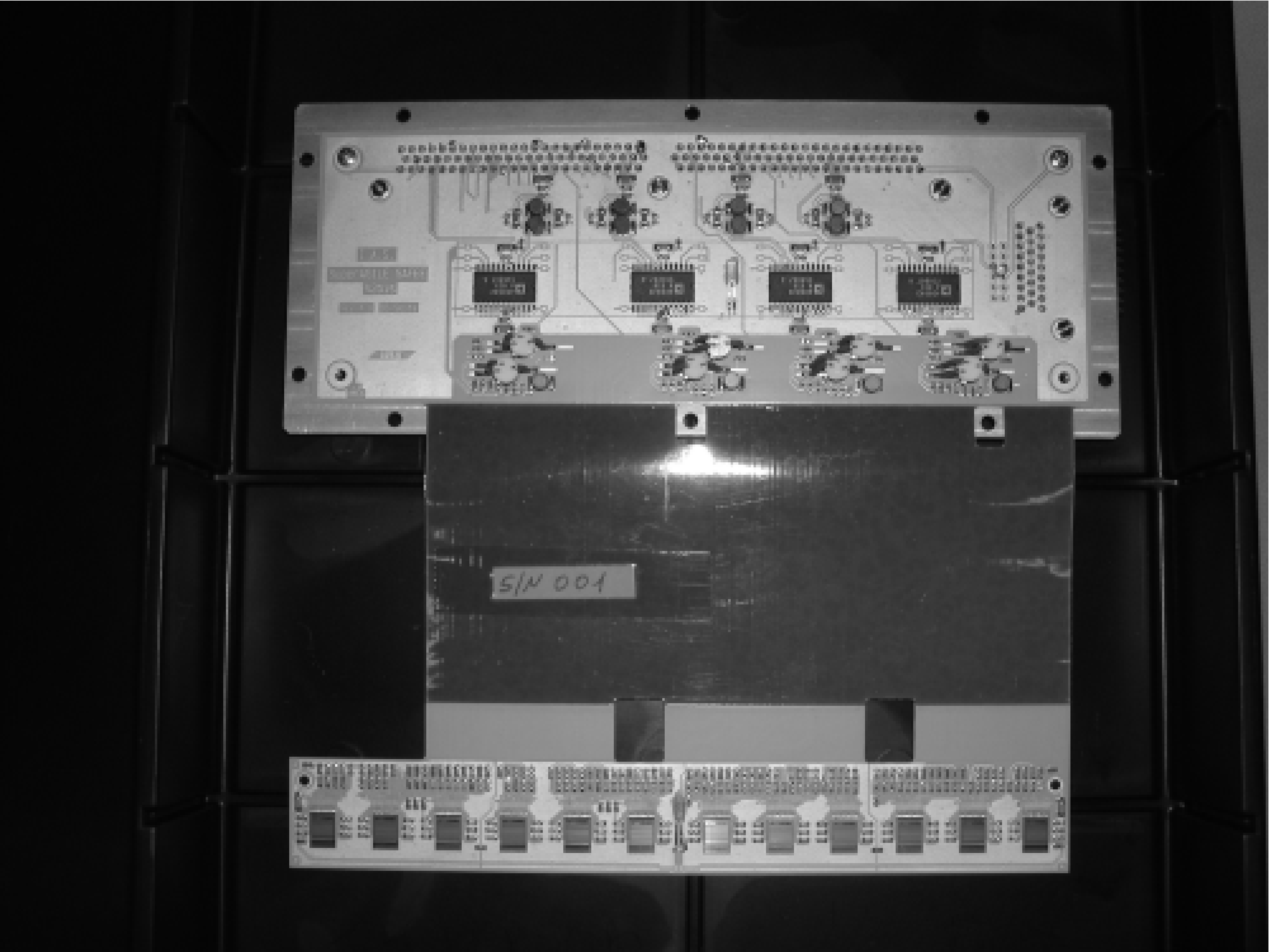}
\caption{Picture of a SAFEE board before the integration with the
detector: the horizontal section (with the XAA1.2 chips) is at the
bottom and the vertical section (with the conditioning electronics)
is on top.} \label{fig:SAFEE_photo}

\end{figure}

At 14.1 cm above the detector the coded mask is fixed, manufactured
on a single tungsten plate 120 $\mu$m thick, glued on a carbon fiber
layer 500 $\mu$m thick. The mask plate is subdivided into four one
dimension modules, each one of the same surface as the detector
module below and with 787 pixels of 19 cm length and 242 $\mu$m
width. The mask-detector geometry gives a resulting pixel size of 6
arcmin. For intense sources the location accuracy can be improved as
the inverse of the signal to noise ration (S/N), down to a
systematics expected of about 1 arcmin. The coded mask is supported
by a tungsten (100 $\mu$m) and carbon fiber (500 $\mu$m) collimator,
that defines also the instrument field of view as two orthogonal
$107 \degree \times 68 \degree$. A picture of the SuperAGILE flight
model is shown in fig. \ref{fig:SuperAGILE_photo}.

\begin{figure}[h!]\centering
\includegraphics[width=14 cm]{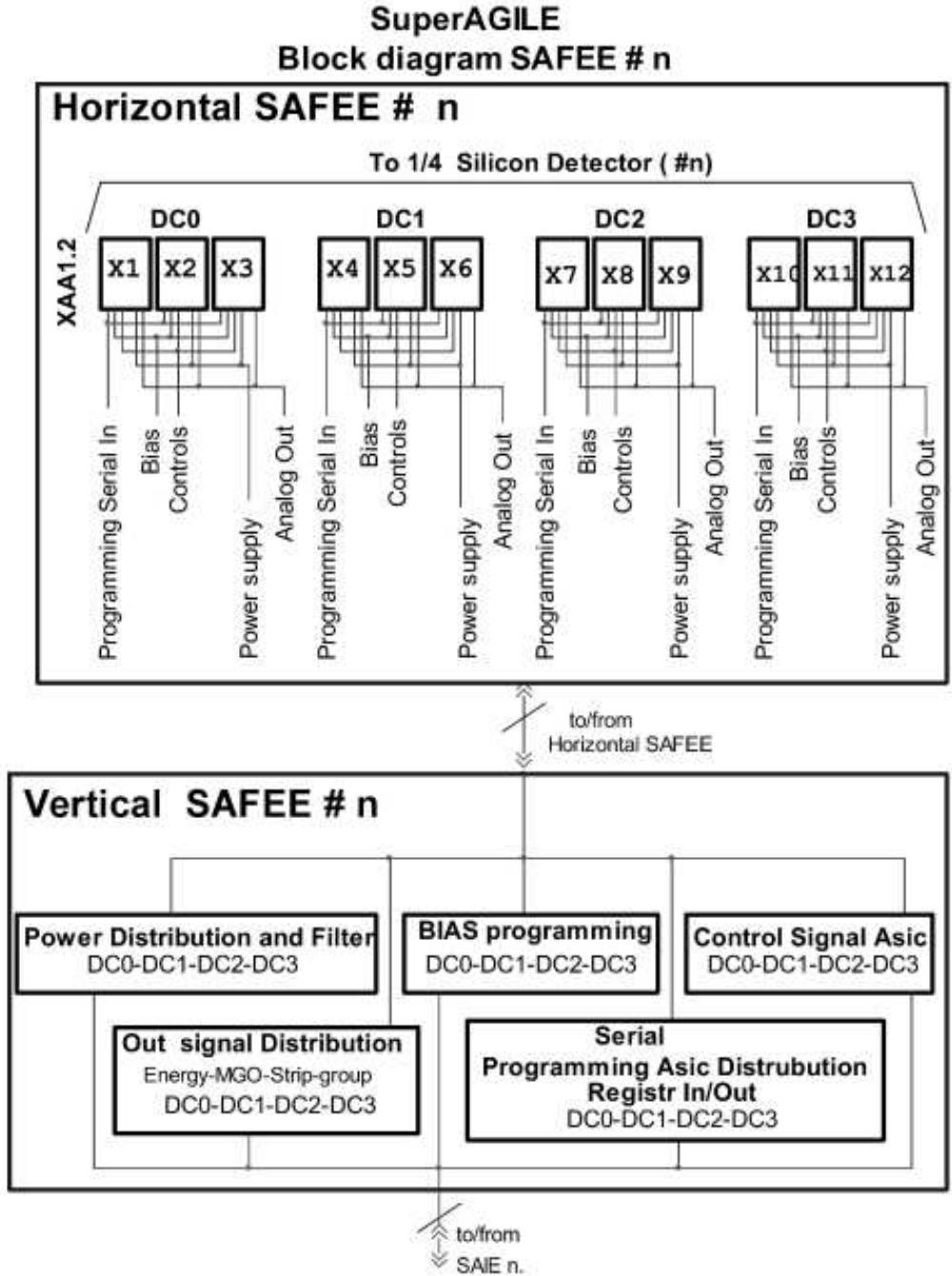}
\caption{Block diagram of the SuperAGILE Front-end Electronics
(SAFEE). The horizontal section contains the daisy chains of the
XAA1.2 front-end electronics chips. The vertical section contains
the power supply distribution, filters, bias generation, serial
programming and output buffers.}\label{fig:SAFEE_block}
\end{figure}

The SuperAGILE front-end electronics (SAFEE) is composed of four
Rigid-Flex Printed Circuit Boards (PCB), each one containing twelve
chips XAA1.2 and interfaced to a detector module using pitch
adapters. In the SAFEE layout the XAA1.2 are arranged in four daisy
chains of three chips each and a daisy chain is considered as a
single and independent unit with its input and output connections.
The power supply, biases (voltages and currents) and output bus are
common in a single daisy chain. The biases and reference voltage to
a single daisy chain are generated using an octal Digital to
Analogue Converter (DAC) chip, AD8842AR unit. Because of the small
room on the SuperAGILE tray, the SAFEE board is divided in two
sections: the horizontal SAFEE, lying on the the detection plane and
containing the XAA1.2 chips, and the vertical SAFEE, containing the
DAC units and all the other conditioning electronics (power supply
distribution, filters, bias generation, serial programming and
output buffers) and fixed on the collimator external wall. A picture
of the SAFEE PCB before the integration to the detector is shown in
fig. \ref{fig:SAFEE_photo}. Two AD590KF temperature sensors are
mounted on the SAFEE PCB, one in the horizontal board and the other
in the vertical board. A block diagram of the SAFEE electronics is
shown in fig. \ref{fig:SAFEE_block}.

\section{Description of the XAA1.2 chip}
\label{sec:XA_features}

\begin{figure}[thp]\centering
\includegraphics[width=15 cm]{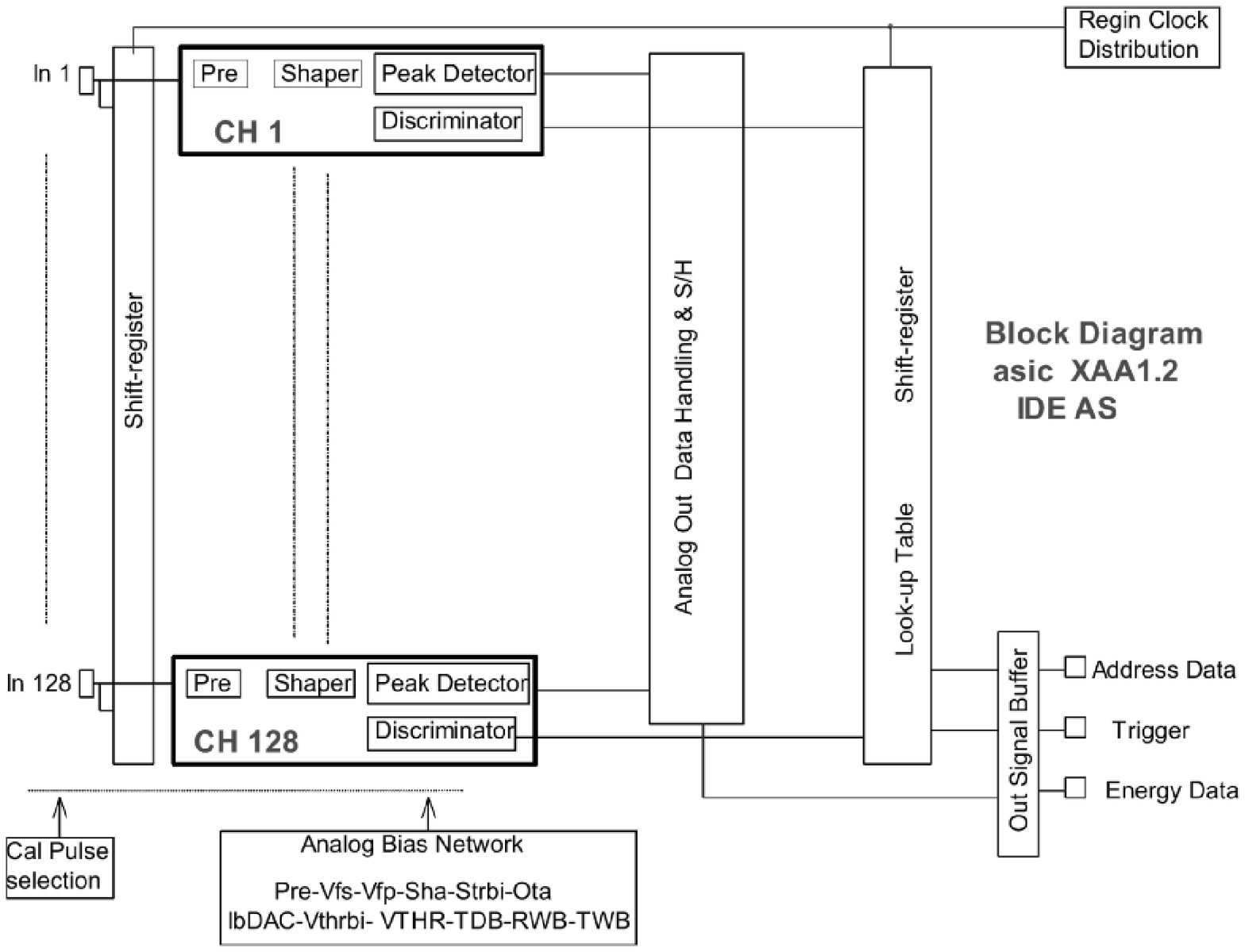}
\caption{Block diagram of the XAA1.2 circuit. All the 128 channels
contain pre-amplifier, shaper, peak detector with stretcher and
discriminator. The digital section is shared among the channels.
From the output buffer the energy, address and trigger (with
multiplicity) signals are provided. A schematic of the XAA1.2
analogue channel can be found in
\cite{Del_Monte_et_al_2005}.}\label{fig:XAA1.2_analog}
\end{figure}

The XAA1.2 is an application specific integrated circuit (ASIC)
designed by means of the very large scale integration (VLSI)
technology by Ideas ASA (now Gamma Medica-Ideas\footnote{Web site:
\texttt{www.gammamedica.com}}) as a front-end electronic chip for
silicon microstrip detectors. The chip is manufactured with a 0.8
$\mu$m complementary metal oxide semiconductor (CMOS) double-poly
and double metal technology (corresponding to a 16 nm gate-oxide
thickness) on epitaxial layer. A block diagram of the XAA1.2,
showing the analog and digital sections of the circuit, can be seen
in fig. \ref{fig:XAA1.2_analog}. The XAA1.2 has been partially
optimized as front-end electronics of the SuperAGILE instrument. In
particular, the input FETs were modified in order to match the 30
$\mathrm{pF \cdot strip^{-1}}$ load of the SuperAGILE detectors and
the power consumption lowered of a factor 3--6.

The XAA1.2 surface is $6 \times 8 \; \mathrm{mm^2}$ and its
thickness is 600 $\mu$m. The circuit is divided into 128 data driven
and self-triggering input channels, consisting of an analogue and a
digital section, and is designed to detect single hit events with
sparse readout. Each analogue signal processing chain is in turn
composed of a charge sensitive preamplifier, a CR-RC shaper and a
peak sensitive stretcher, as sketched in fig.
\ref{fig:XAA1.2_analog}. The peak stretcher can be excluded from the
analogue processing of one circuit by changing the digital
configuration of the ASIC, thus reducing the overall power
consumption. The analogue chains are followed by a Analogue Out Data
Handling and Sample and Hold (S/H) stage. The main features of the
XAA1.2 as a front-end electronic circuit for silicon microstrip
detectors were reported in \cite{Soffitta_et_al_2000}. The XAA1.2 is
a custom ASIC chip designed starting from the commercial device
XA1.3 aiming to improve the performance in terms of noise, power
consumption and thermal stability after the extensive laboratory
testing reported in \cite{Del_Monte_et_al_2000}.

The XAA1.2 digital section can be programmed by serially loading an
internal 646-bit register. The register content (referred to as
configuration mask or ``Register In'') encodes all the information
needed to configure the chip: 128 bit specifying the enabled
channels in test mode (for an electrical stimulus to be injected as
described in section \ref{sec:linearity_noise_ecal}), 384 bit
configuring the threshold fine regulation (3 bit per channel), 128
bit specifying the disabled channels, 4 bit containing the address
offset (to address up to sixteen XAA1.2 connected in daisy chain), 1
bit enabling the stretcher and 1 bit enabling the chip in test mode.
By sending twice the configuration mask the actual value contained
in the memory (referred to as ``Register Out'') is provided and can
be read out.

The XAA1.2 is supplied using four highly regulated voltages, two for
the analogue section ($\mathrm{AV_{DD}=+2.0 \; V}$ and
$\mathrm{AV_{SS}=-2.0 \; V}$) and two for the digital section
($\mathrm{DV_{DD}=+2.0 \; V}$ and $\mathrm{DV_{SS}=-2.0 \; V}$). The
power consumption of the digital part is negligible if compared with
the analogue section. The measured supply currents of the chip in
nominal bias conditions (shown in tab. \ref{tab:bias}) and
stretcher-on configuration are $\mathrm{I_{DD} \simeq 29 \; mA}$ and
$\mathrm{I_{SS} \simeq 38 \; mA}$, thus giving an overall power
consumption of about 1 $\mathrm{mW \cdot ch.^{-1}}$.

The XAA1.2 functionality and performance are controlled by eleven
biases and a reference voltage, all provided to the chip using
external DAC units. The bias currents are \textit{prebias} and
\textit{shabias} (controlling the source and the drain current of
the preamplifier and shaper respectively), \textit{otabias} (bias
current of the discriminator), \textit{IbDAC} (controlling the
voltage to current buffer converters), \textit{strbias} (bias
current of the peak stretcher), \textit{Vthrbi} (controlling the
digital fine regulation of the threshold), \textit{TrigDelbias}
(setting the delay between the trigger and the internal sample and
hold), \textit{TrigWbias} (defining the hold width) and finally the
\textit{ResWbias} (affecting the width of the reset signal after an
event is detected and processed). The bias voltages are \textit{Vfs}
and \textit{Vfp} (DC voltages for the gate of the feedback MOS
resistor in the preamplifier and the shaper respectively) and the
reference voltage is \textit{Vthr}. The default value of the bias
currents and voltages are shown in tab. \ref{tab:bias}.

\begin{table}[h!]
\centering \caption{XAA1.2 typical bias currents and voltages.}
\bigskip
\label{tab:bias}

\begin{tabular}{c @{$\qquad\qquad\qquad\qquad$} c}

Bias name            & bias value \\

\hline

\textit{prebias}     & 80  $\mu$A \\
\textit{shabias}     &  9  $\mu$A \\
\textit{otabias}     & 10  $\mu$A \\
\textit{strbias}     & 60  $\mu$A \\
\textit{Vthrbi}      & 20  $\mu$A \\
\textit{IbDAC}       & 60  $\mu$A \\
\textit{TrigDelbias} &  2  $\mu$A \\
\textit{TrigWbias}   & 0.2 $\mu$A \\
\textit{ResWbias}    & 0.1 $\mu$A \\
\textit{Vfp}         & -125 mV    \\
\textit{Vfs}         & 960 mV \\

\end{tabular}
\end{table}

The XAA1.2 is developed as a front-end electronics chip for position
sensitive silicon microstrip detectors. Following the internal chip
logic the 128 channels are divided into four groups of 32 channels
each. For this reason a single channel is identified by its position
within the group (\textit{strip} signal, 5 bit) and of its group
within the chip (\textit{group} signal, 2 bit). To the latter signal
4 offset bit can be externally added, encoded in the Register-In, to
allow the identification of the chip in a daisy chain (of 16 XAA1.2
maximum). Since the internal chip logic is not able to provide
information on more than one hit at a time, the switch of a
threshold inhibits all the other inputs within a few nanoseconds
range. If more than one event is present during this time interval,
a multiple event is detected and its output signals are meaningless.
The multiplicity of the processed events is provided by a
multiplicity generator output (\textit{MGO} signal), thus allowing
to reject multiple hit events. Internal DAC units convert the
\textit{strip}, \textit{group} and \textit{MGO} signals into
analogue current differential outputs, in order to reduce the
pick-up noise. Consequently on trigger the XAA1.2 returns four pairs
of differential output current signals: amplitude, \textit{strip},
\textit{group} and \textit{MGO}.

The amplitude output signal from the ASIC shaper is affected by
dumped oscillations, lasting for about 30 $\mu$s. This ``wiggle'' is
a constitutional property of the XAA1.2, due to a non perfectly
matched mirroring of the currents inside the chip (B. Sundal,
private communication), and cannot be fixed from the exterior. The
``wiggle'' does not allow to sample the signal before 20 $\mu$s from
the trigger and is a lower limit to the XAA1.2 dead time.

The analogue chains of the XAA1.2 can process charge pulses produced
by an external generator, that is used to calibrate the electronics
by replicating the output pulses from a detector. For this purpose
the chip contains a test charge input where voltage pulses can be
provided. Such voltage signals are then converted into charge by an
internal 0.5 pF capacitor.

The threshold is set to the XAA1.2 using an external reference
voltage, thus giving the same reference value for all the input
channels. Moreover the XAA1.2 channels allow a 3-bit fine regulation
of the analogue threshold, encoded in the configuration mask, to
reduce the threshold intrinsic disuniformity.

\section{Experimental set-up}
\label{sec:set-up}

The measurements reported in this paper have been performed with the
XAA1.2 as part of the SAFEE circuit. For testing purposes, the SAFEE
PCB is connected to a dedicated test equipment, based on VME bus
with VME-PCI bridge and working under Linux operating system. The
test equipment provides the power supply, programs the XAA1.2
internal registers, generates the voltage pulse to feed the test
charge input, converts the output current signals to voltage,
performs the analogue to digital conversion using a 12-bit Analogue
to Digital Converter (ADC) and adds the time tag to the events.
Particular care was devoted to the control and reduction of the
electronic noise when designing, building and operating the test
equipment. A detailed description may be found in
\cite{Pacciani_et_al_2006}. Further details about the experimental
set-up are given in the sections dedicated to the measurements.

\section{Analytic estimation of the XAA1.2 noise contributions}
\label{sec:noise_th}

The main sources of noise in ASIC circuits coupled to silicon
microstrip detectors are the shot noise given by the leakage current
in the detector, the thermal noise from the detector bias resistors,
the noise produced by the metal strips and the transistor channel
noise in the readout chip. All the other noise components in the
input transistor, such as the flicker noise and the bulk-series
resistance noise, are negligible when compared to the contributions
listed above. In our set-up the channel noise is by far dominant
over the other contributions. In this section an overview on the
main sources of noise is given and each contribution is specialized
to the case of the XAA1.2 chip, following the method in
\cite{Barichello_et_al_1998} applied to the silicon microstrip
detector described in sec. \ref{sec:SuperAGILE}.

The noise figures are generally expressed using the Equivalent Noise
Charge ($ENC$), a parameter connecting the root mean squared ($RMS$)
noise voltage at the output of the shaper with the signal input and
generally measured in $\mathrm{e^-}$. All the noise sources increase
the width of a monochromatic energy input following a gaussian
distribution. For this reason, the $ENC$ (in $\mathrm{e^-}$) can be
converted into the Full Width at Half Maximum ($FWHM$) of a spectral
line (in keV) taking into account the equation connecting the $FWHM$
and the signal $RMS$:

\begin{equation}\label{eq:ENC_FWHM}
    FWHM \simeq \frac{2.35 \cdot 3.62 \cdot ENC}{1000}.
\end{equation}

In case of the XAA1.2 chip, the shot noise given by the leakage
current in the detector can be written as

\begin{equation}\label{eq:ENC_leak}
    ENC_{leak}=107 \cdot \sqrt{I_L(T) \cdot t_P}
\end{equation}

\noindent where $I_{L}$ is the strip leakage current (in $\mathrm{nA
\cdot cm^{-2}}$), $t_P$ is the peaking time (in $\mu$s) and $T$ is
the temperature (in K). Considering that at room temperature (300 K)
the SuperAGILE detector leakage current is $I_{L}=2 \; \mathrm{nA
\cdot cm^{-2}}$ and the signal peaking time is $t_P=2 \; \mu$s, the
contribution is $ENC_{leak} \simeq 200 \; \mathrm{e^-}$ and
$FWHM_{leak} \simeq 1.8$ keV.

The thermal noise in the detector bias resistors can be written as

\begin{equation}\label{eq:ENC_bias}
    ENC_{bias}=45 \cdot \sqrt{\frac{t_P \cdot T}{R_{bias}}}
\end{equation}

\noindent with $R_{bias}=20 \; M \Omega$. Substituting the numeric
values applicable for SuperAGILE the estimated noise is $ENC_{bias}
\simeq 240 \; \mathrm{e^-}$ and $FWHM_{bias} \simeq 2.0$ keV.

The ENC contribution of the detector metal strips is

\begin{equation}\label{eq:ENC_ms}
    ENC_{ms}=2.58 \cdot 10^{-2} \cdot C_{IN} \cdot \sqrt{\frac{R_{ms} \cdot T}{t_P}}
\end{equation}

\noindent and since for the SuperAGILE silicon microstrip detector
$C_{IN}=30 \; \mathrm{pF \cdot strip^{-1}}$ and $R_{ms}=72 \;
\Omega$, $ENC_{ms} \simeq 80 \; \mathrm{e^-}$ and $FWHM_{ms} \simeq
0.7$ keV.

The XAA1.2 transistor channel noise is in principle given by

\begin{equation}\label{eq:ENC_el1}
    ENC_{XAA1.2}=ENC_{1/f} \oplus ENC_{ch} \oplus ENC_{bulk}
\end{equation}

\noindent where $ENC_{1/f}$ is the flicker noise contribution,
$ENC_{ch}$ is the transistor channel component and $ENC_{bulk}$ is
the transistor bulk-resistance noise. As shown in
\cite{Barichello_et_al_1998}, the contributions from flicker and
bulk resistance noise can be neglected and the transistor channel
noise at the \textit{prebias} nominal value of $50 \; \mu A$ can be
written with the semiempirical formula

\begin{equation}\label{eq:ENC_el}
    ENC_{XAA1.2}=171+23.5 \cdot C_{IN}
\end{equation}

\noindent composed of a contribution proportional to the input
capacitance and a constant term taking into account the internal
capacitance. The parameters in the (\ref{eq:ENC_el}) have been
estimated from the laboratory measurements with the test pulse
generator and include the Johnson noise contribution. Taking into
account the capacitance of the detector ($C_{IN}=30$ $\mathrm{pF
\cdot strip^{-1}}$) the resulting noise is $ENC_{XAA1.2} \simeq 880
\; \mathrm{e^-}$ and $FWHM_{XAA1.2} \simeq 7.5$ keV.

Considering the mean energy needed to create a hole-electron pair in
silicon ($E_0=3.62$ eV), the fluctuation on the number of charge
carriers produced by the interacting photon is one of the noise
contributions introduced by the detector. Taking into account that
the number of electron-hole pairs in the detector follows the
Poisson statistics (corrected by the Fano factor) and that only the
holes are processed by the XAA1.2, the contribution is

\begin{equation}\label{eq:ENC_det}
    ENC_{stat}=\sqrt{N_h \cdot f}
\end{equation}

\noindent where $N_h$ is the number of detected holes and $f$ is the
Fano factor of the material. Since in silicon $f=0.14$ (as reported
for example in \cite{Knoll_1989}), typical values of the detector
noise contribution are $ENC_{stat} \simeq 20 \; \mathrm{e^-}$
(corresponding to $FWHM_{stat} \simeq 0.2$ keV) at 10 keV,
$ENC_{stat} \simeq 44 \; \mathrm{e^-}$ (corresponding to
$FWHM_{stat} \simeq 0.4$ keV) at 50 keV and $ENC_{stat} \simeq 62 \;
\mathrm{e^-}$ (corresponding to $FWHM_{stat} \simeq 0.5$ keV) at 100
keV.

The total noise in the XAA1.2 chip is given by the combination of
the contributions discussed above,

\begin{equation}\label{eq:ENC_tot}
    ENC_{total}=ENC_{stat} \oplus ENC_{leak} \oplus ENC_{bias} \oplus ENC_{ms} \oplus ENC_{XAA1.2}
\end{equation}

\noindent and it is found in this section that the transistor
channel noise is by far the most important noise source. For
SuperAGILE detector and front-end electronics the resulting $ENC$
from the (\ref{eq:ENC_tot}) is $ENC_{total} \simeq 935 \;
\mathrm{e^-}$ and $FWHM_{total} \simeq 8.0$ keV.

\section{Measurement of the XAA1.2 linearity and noise using the charge test pulse generator}
\label{sec:linearity_noise_ecal}

The linearity of all the XAA1.2 input channels can be measured using
the test pulse generator (described in sec. \ref{sec:XA_features}).
For each channel the mean value of the digitized output amplitude as
a function of the input charge represents a measure of the linearity
(referred to as ``calibration curve''). Applying a linear fit to the
calibration curve, the linearity parameters of the channels are
separately evaluated in terms of offset (intercept of the linear
fit) and gain (angular coefficient of the linear fit). A sequential
measure of the linearity on all the XAA1.2 input channels with the
test pulse generator is referred to as ``electronic calibration''.
All the measurements reported in this section deal with the XAA1.2
connected to the $30 \; \mathrm{pF \cdot strip^{-1}}$ silicon
microstrip detector through wire bondings.

\begin{figure}[hp!]\centering

\includegraphics[width=10 cm, angle=90]{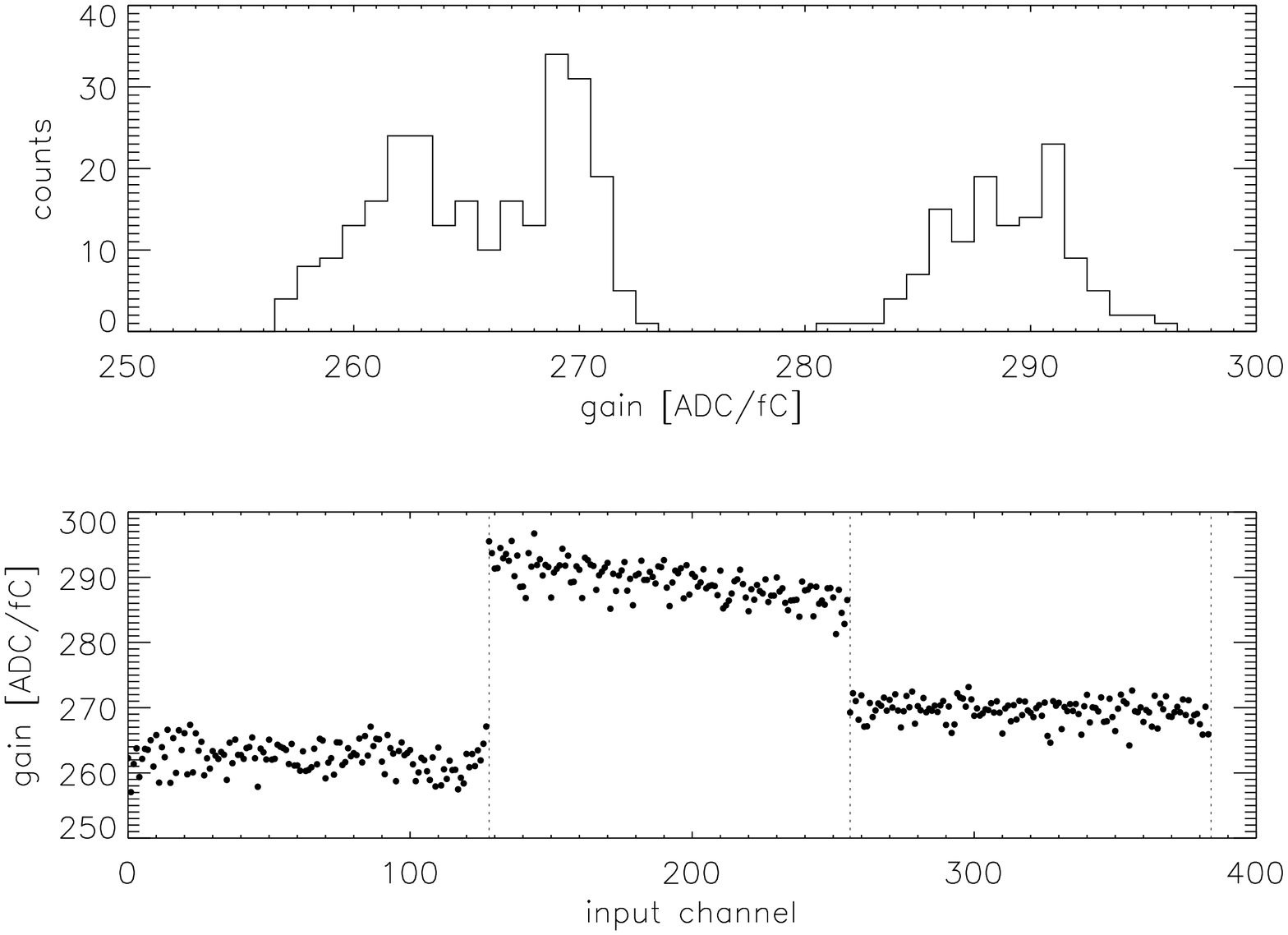}
\caption{Histogram of the gain (upper panel) and gain for each strip
in a daisy chain after bonding the XAA1.2 chip to the silicon
microstrip detector. The vertical dotted lines in the lower panel
separate the three chips within the daisy
chain.}\label{fig:gain_det}

\includegraphics[width=10 cm, angle=90]{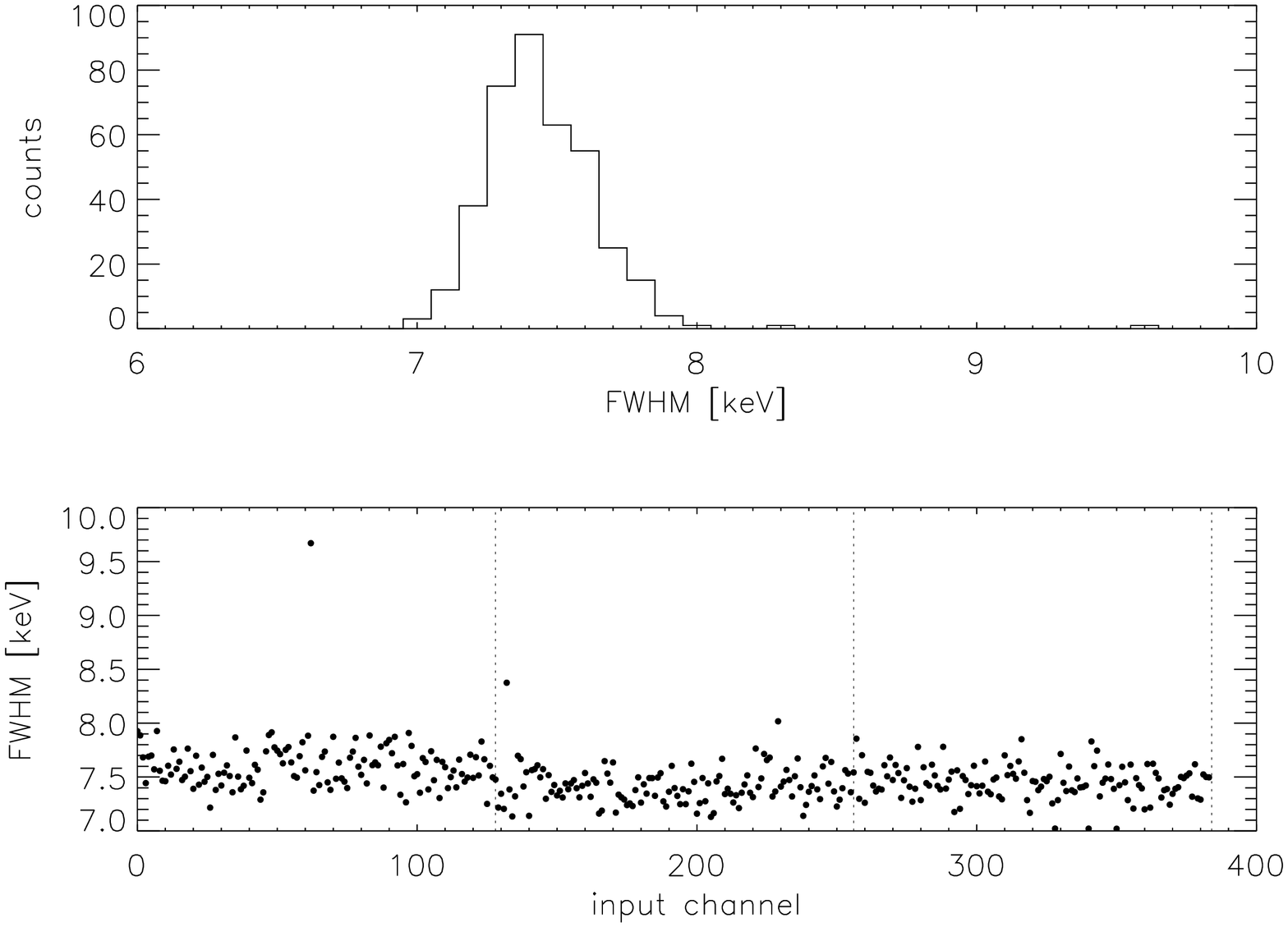}
\caption{Histogram of the noise FWHM (upper panel) and noise FWHM
for each strip in a daisy chain after bonding the XAA1.2 chip to the
silicon microstrip detector. The vertical dotted lines in the lower
panel separate the three chips within the daisy
chain.}\label{fig:FWHM_det}

\end{figure}

In the electronic calibration the XAA1.2 is typically fed with input
charges ranging from 0.5 fC (corresponding to about 11 keV in
silicon) up to 5 fC (corresponding to about 113 keV in silicon).
From these measurements we find that the XAA1.2 is linear up to 5
fC, the average residual of the linear fit being $1.3 \sigma$ at
this input charge value. An example of the distribution of the gain
values of the 384 channels in a daisy chain is plotted in fig.
\ref{fig:gain_det}. In the upper panel the histogram of the gain
values is plotted while in the lower panel the gain value of each
input channel is plotted. The vertical dotted lines in the lower
panel separate the three chips within the daisy chain. The
represented data are typical of the XAA1.2 chips, showing their
level of disuniformity and difference.

The XAA1.2 noise is evaluated in each input channel as the mean
value of the FWHM of the output amplitude histogram for the
different input charges. The FWHM value, initially measured in ADC
channels, is then converted into energy units by dividing it by the
gain, obtained from the linearity fit. An example of the
distribution of the FWHM noise values of the same 384 channels (as
in fig. \ref{fig:gain_det}) is plotted in fig. \ref{fig:FWHM_det}.
In the figure the upper panel shows the noise FWHM values histogram
while the lower panel shows the noise FWHM values for each input
channel. The vertical dotted lines in the lower panel separate the
three chips within the daisy chain. The noise FWHM averaged over the
384 input channels is 7.5 keV with a 0.2 keV standard deviation.
Again, the data reported here for three chips are representative of
the general XAA1.2 behaviour. The measured FWHM is smaller than the
theoretical estimations in sec. \ref{sec:noise_th} because the noise
has been reduced adjusting the amplitude signal shape following the
method described in sec. \ref{sec:bias_optimization}.

\section{Measurement of the XAA1.2 linearity and noise using the X-ray sources}
\label{sec:linearity_noise_X-rays}

A verification of the XAA1.2 linearity and noise has been performed
by the acquisition of photons from calibration radioactive X-ray
sources. The sources used in these measurements are
$^{109}\mathrm{Cd}$ (Ag K$\alpha$ line at 22.1 keV),
$^{241}\mathrm{Am}$ (decay line at 59.5 keV), $^{57}\mathrm{Co}$
(decay line at 122.1 keV) and a custom source emitting Ba
fluorescence photons at 32.1 keV from a $\mathrm{BaF_2}$ crystal
excited by two $^{241}\mathrm{Am}$ sources.

The measurement statistics does not allow to fit the spectrum
acquired by a single strip, so the spectra have been accumulated on
a daisy chain. Since each microstrip can be considered as an
independent detector, with its own gain parameters, a correction
algorithm need to be applied to the data in order to accumulate the
spectra. For this reason, the digitized amplitude (in ADC channels)
of each detected photon has been converted into a conventional
amplitude (in PHA) using the gain parameters of the triggered input
channel, evaluated with the test charge measurements as in sec.
\ref{sec:linearity_noise_ecal} under the assumption that the test
capacitor has the same value in each chip.

\begin{figure}[hp!] \centering

\includegraphics[width=10 cm, angle=90]{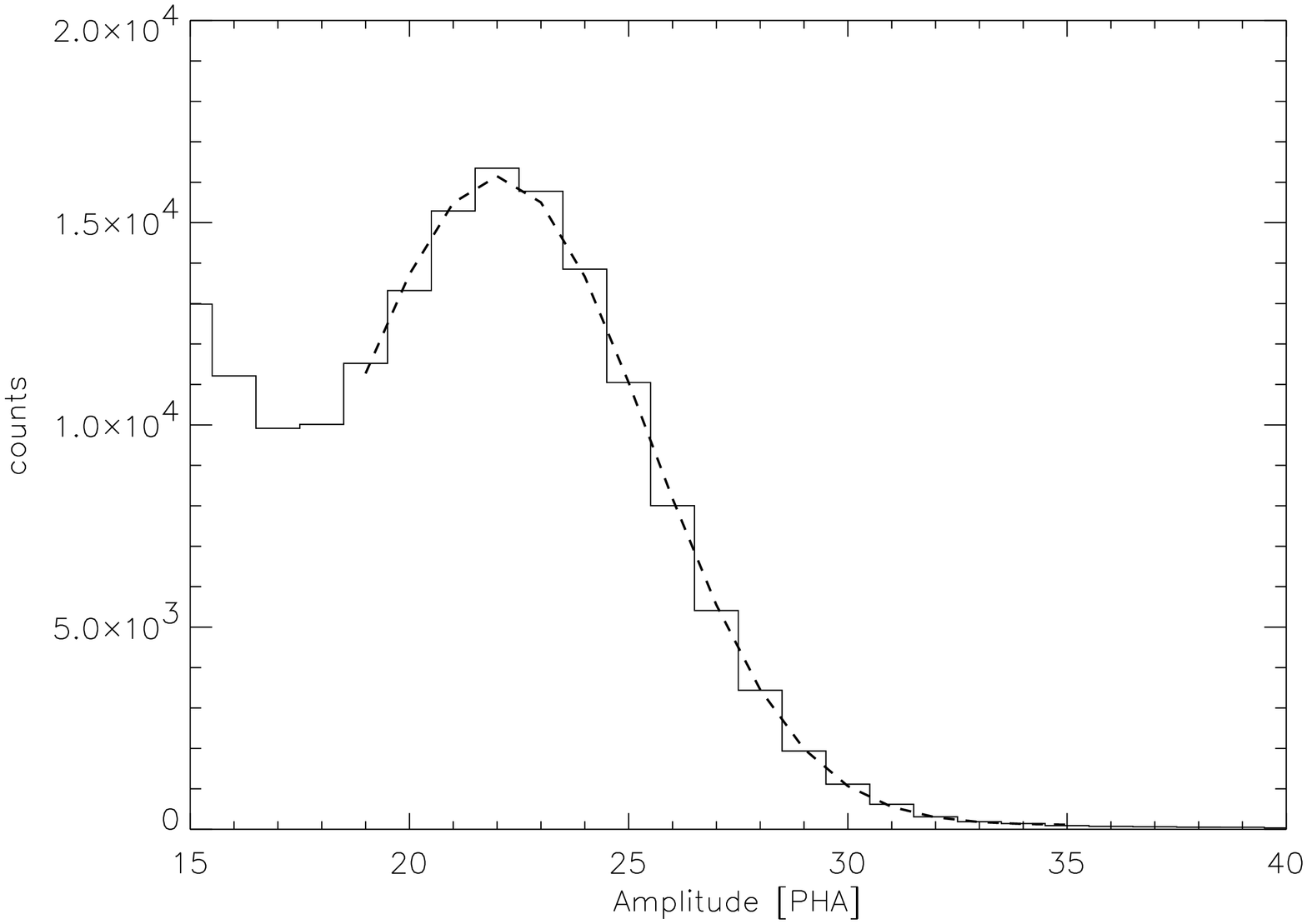}
\caption{Spectrum of the 22.1 keV line of the $^{109}\mathrm{Cd}$
source with the dashed fitting curve superimposed (two gaussian
functions plus a constant background).}\label{fig:Cd109_fit}

\includegraphics[width=10 cm, angle=90]{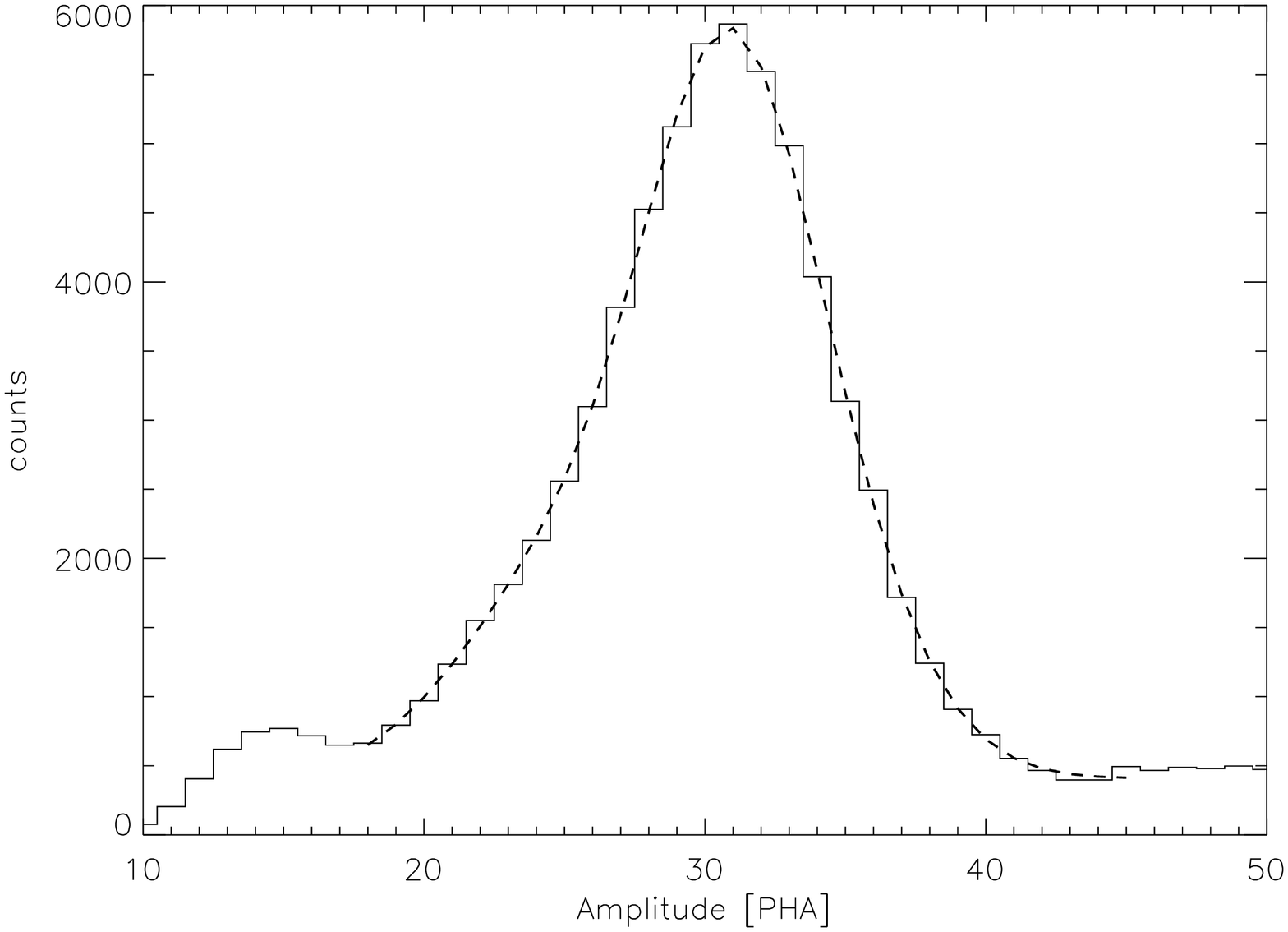}
\caption{Spectrum of the 32.1 keV line of the $\mathrm{BaF_2}$
source with the dashed fitting curve superimposed (two gaussian
functions plus a constant background).}\label{fig:BaF2_fit}

\end{figure}

\begin{figure}[p] \centering

\includegraphics[width=10 cm, angle=90]{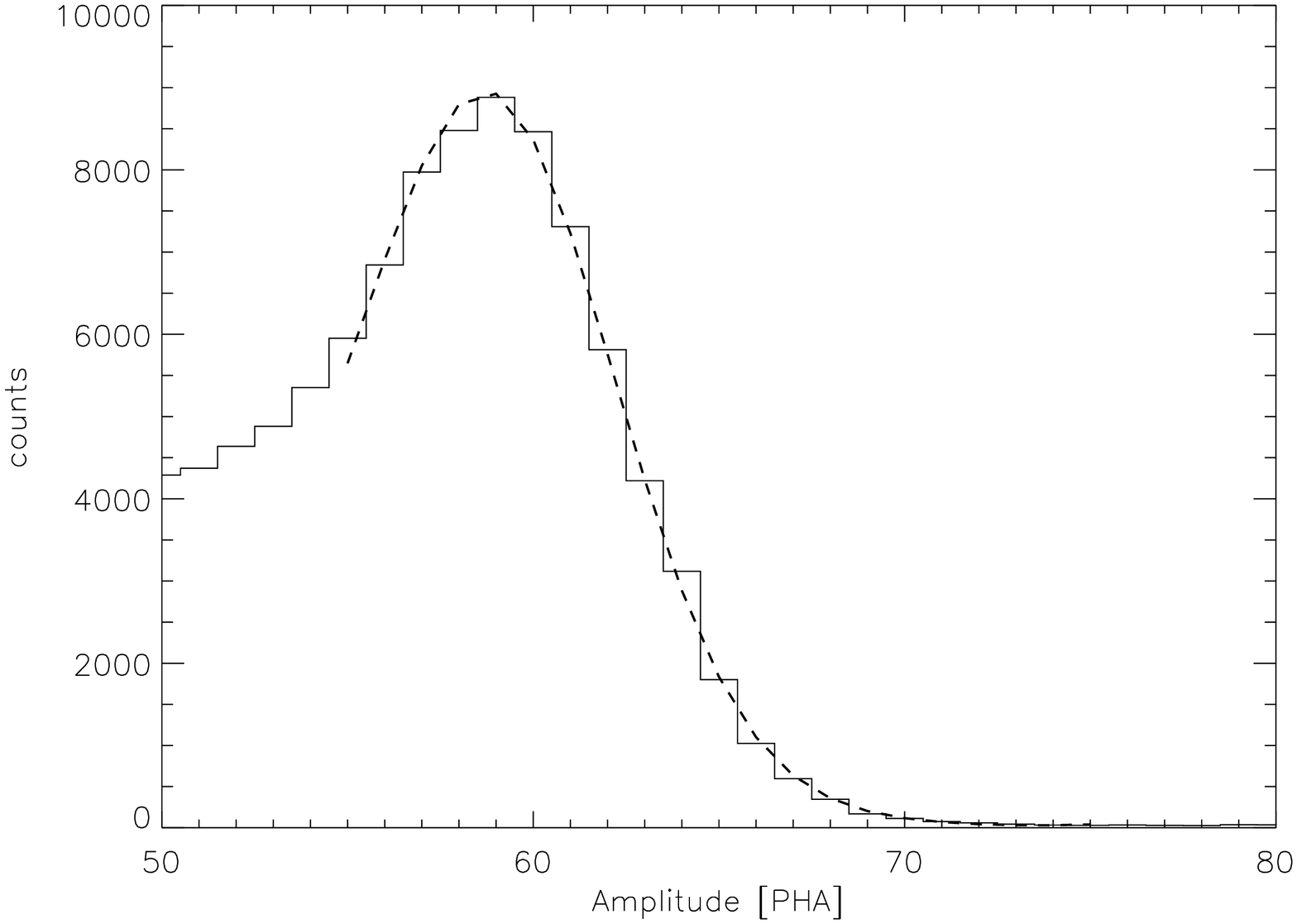}
\caption{Spectrum of the 59.5 keV line of the $^{241}\mathrm{Am}$
source with the dashed fitting curve superimposed (a gaussian
function plus a quadratic background).}\label{fig:Am241_fit}

\includegraphics[width=10 cm, angle=90]{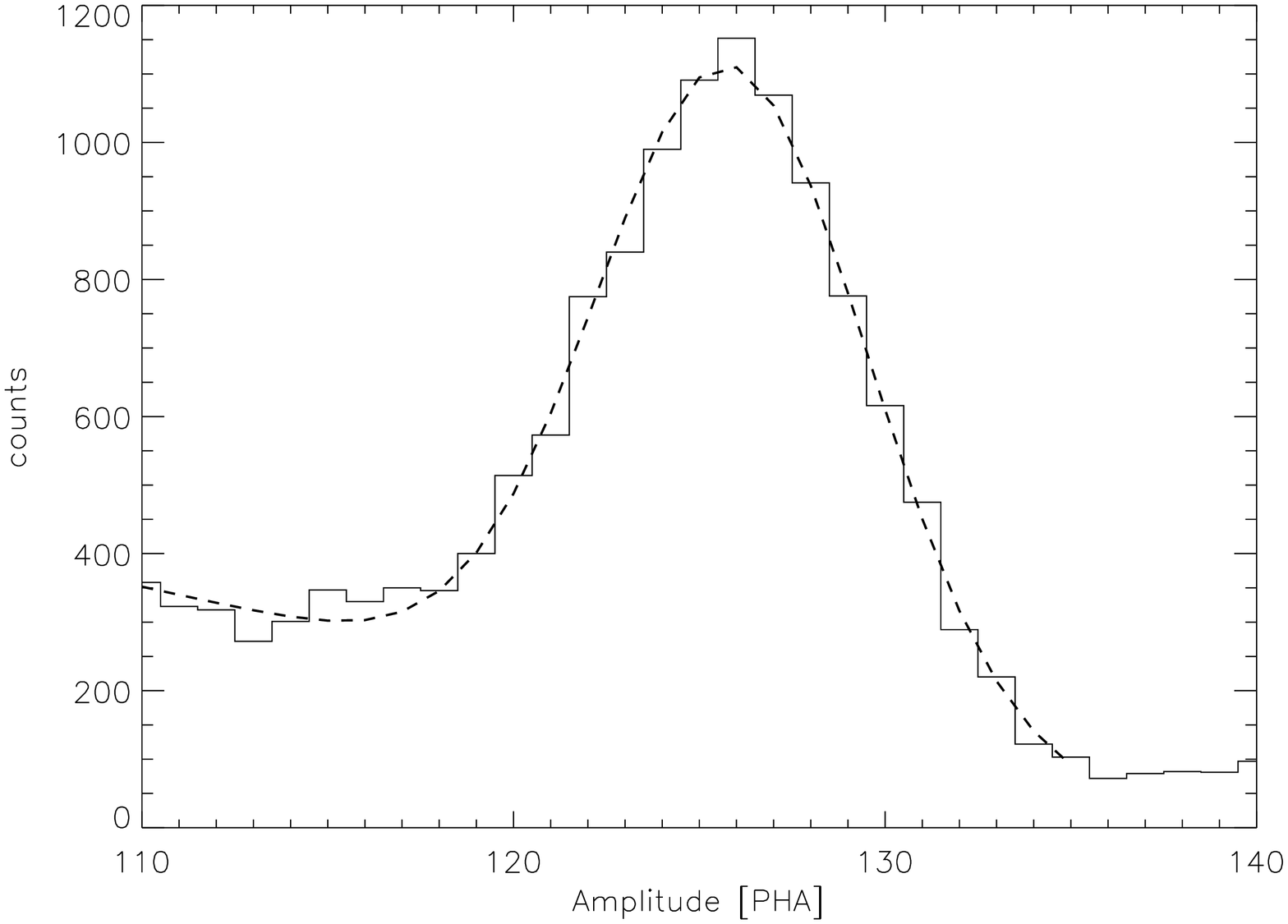}
\caption{Spectrum of the 122.1 keV line of the $^{57}\mathrm{Co}$
source with the dashed fitting curve superimposed (a gaussian
function plus a linear background).}\label{fig:Co57_fit}

\end{figure}

As an example, in fig. \ref{fig:Cd109_fit}, \ref{fig:BaF2_fit},
\ref{fig:Am241_fit} and \ref{fig:Co57_fit} the spectra acquired with
the $^{109}\mathrm{Cd}$ (22.1 keV line), custom $\mathrm{BaF_2}$
(32.1 keV line), $^{241}\mathrm{Am}$ (59.5 keV line) and
$^{57}\mathrm{Co}$ (122.1 keV line) are plotted. In order to obtain
a significant statistics, the spectra of each source are accumulated
on 383 input channels from the same subset as in sec.
\ref{sec:linearity_noise_ecal}, apart from the last channel that is
excluded from the acquisition since its noise contribution is much
higher than the other channels because of a lower threshold.

\begin{figure}[p] \centering

\includegraphics[width=10 cm, angle=90]{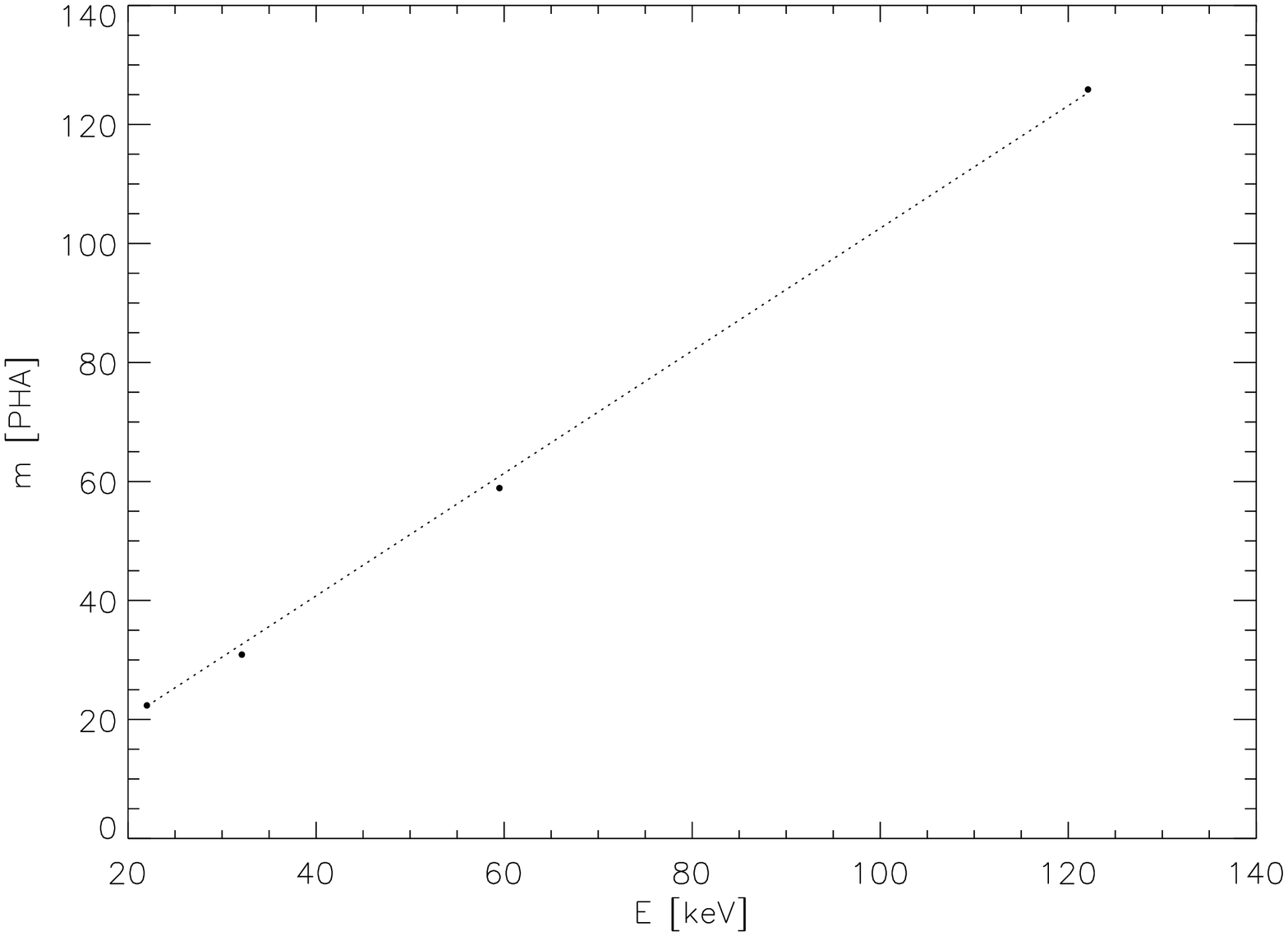}
\caption{Gaussian peak position $m$ as a function of the input X-ray
line energy $E$ with the dotted linear fit superimposed.}
\label{fig:m_vs_E}

\includegraphics[width=10 cm, angle=90]{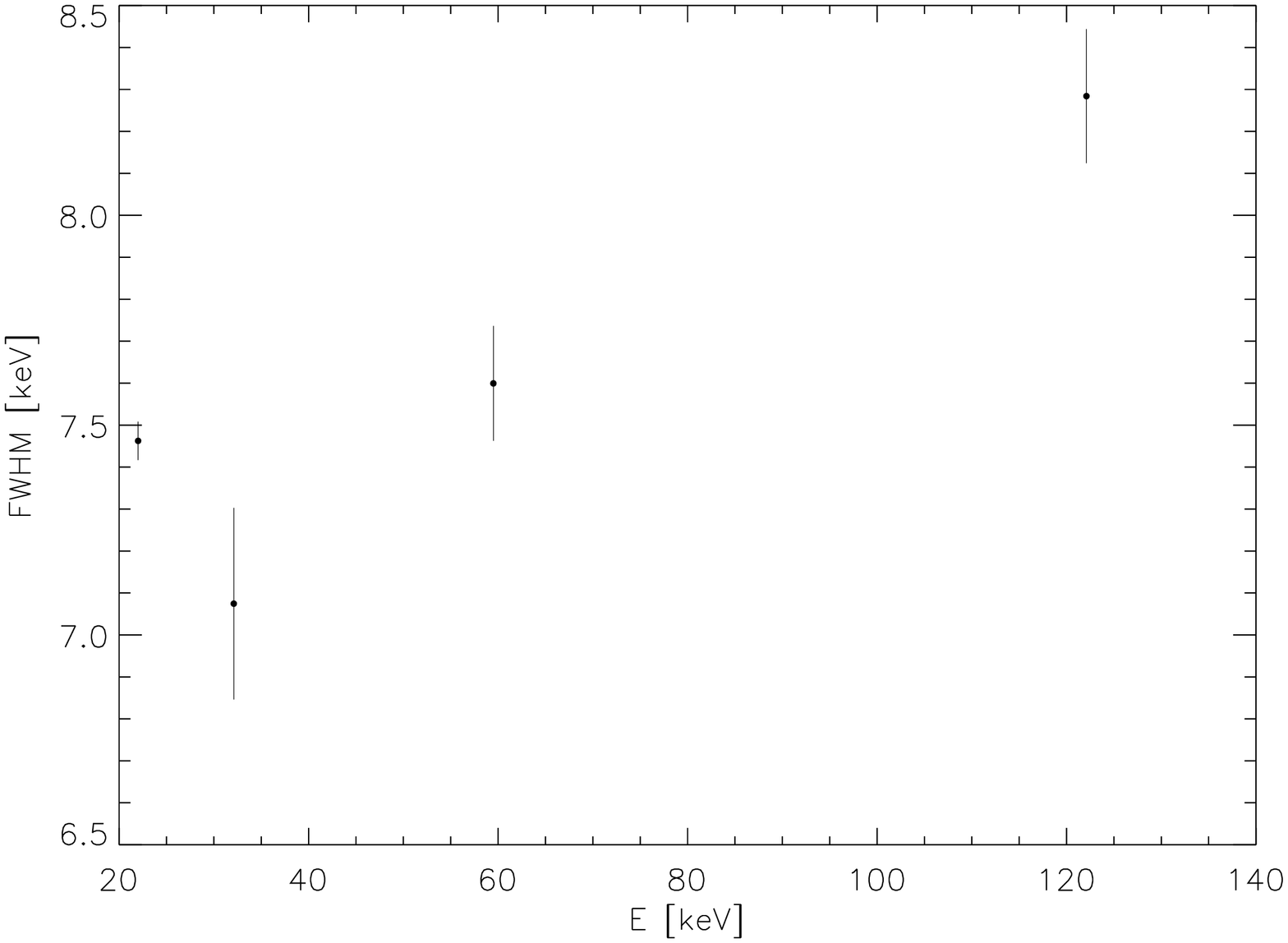}
\caption{Gaussian $FWHM$ as a function of the input X-ray line
energy $E$.}\label{fig:FWHM_vs_E}

\end{figure}

The spectra in the figures from \ref{fig:Cd109_fit} to
\ref{fig:Co57_fit} are fitted with a model based on a gaussian
function representing the X-ray line and a polynomial background
depending on the specific source. In case of the $^{109}\mathrm{Cd}$
and $\mathrm{BaF_2}$ sources, where two lines are present that are
not completely resolved because of the system spectral resolution,
the fitting function includes both the lines, modeled using
gaussians. The peak position $m$ of the gaussian from the fit is
then linearly fitted as a function of the X-ray line energy $E$

\begin{equation}\label{eq:m_vs_E}
  m=a + b \cdot E
\end{equation}

as shown in fig. \ref{fig:m_vs_E}. The parameters of the linear fit
in (\ref{eq:m_vs_E}) applied to the data sample in fig.
\ref{fig:m_vs_E} are $a=-0.41 \pm 0.03$ PHA and $b=1.0298 \pm
0.0006$ $\mathrm{PHA \cdot keV^{-1}}$ and the linear fit is shown as
a dotted line in fig. \ref{fig:m_vs_E}.

The $b$ parameter of the linear fit in (\ref{eq:m_vs_E}) allows to
correlate the linearity measurements performed with the electronic test
pulse and with the X-ray sources, since the amplitude of the
acquired photons is corrected on the basis of the calibration curve of
the triggering channel. The $+3 \; \%$ variation of the fitted $b$
parameter with respect to the expected value of one is a measure
of the systematic error of the correction procedure and is due to
the variation of the input capacitance, that generates the input
test pulse, with respect to the nominal 0.5 pF value. For this
reason, the linear fit in (\ref{eq:m_vs_E}) can be used to
calibrate the test charge generator by correcting the input charge
values in the calibration curve by the $b$ factor.

The noise in the spectrum is estimated from the gaussian standard
deviation $\sigma$ of the spectral fit, converted into the line
$FWHM$ following

\begin{equation}\label{eq:FWHM_vs_sigma}
  FWHM=2.35 \cdot \frac{\sigma}{b}
\end{equation}

where $b$ is the conversion factor between amplitude (PHA) and
energy (keV) units as estimated in (\ref{eq:m_vs_E}). The FWHM
values of the data sample, estimated from (\ref{eq:FWHM_vs_sigma}),
are plotted as a function of energy in fig. \ref{fig:FWHM_vs_E}. As
can be seen in the plot, FWHM from 7.1 keV (at 22.1 keV) up to 8.3
keV (at 122.1 keV) are found and the values are compatible at
$4\sigma$ with the FWHM measured in sec.
\ref{sec:linearity_noise_ecal} using the test pulse generator.

\section{Measurement of the XAA1.2 threshold}
\label{sec:threshold}

In the SAFEE architecture the same reference voltage \textit{Vthr}
is set to the discriminator of all the input channels in a daisy
chain using the external DAC (AD8842AR). During the measurements the
issue of converting the reference voltage into charge (and then into
energy) is raised. In order to have the same operational threshold
definition for both on-ground and in-orbit conditions, where the
noise is different, the SuperAGILE threshold is defined starting
from the background counting rate.

As a first step, background acquisitions are performed at high
reference voltage, where the counting rate is dominated by the
environmental background and the electronic noise contribution can
be neglected. This measurement provides an estimate of the local
environmental background, that depends on the site and will not be
present in orbit. In orbit we expect two major components for the
background counting rate: the Diffuse X-ray Background (DXB, see
\cite{Zombeck} for further information) impinging on the detector
through the open slits of the mask and within the collimator
aperture and the tail of the electronic noise, since the particle
background is very efficiently eliminated by the upper threshold of
the SAIE. Our current operational definition of threshold is such
that the background counting rate is due to the DXB plus a 10 \% due
to electronic noise.

The threshold voltage estimated with the method outlined above is
then converted into a charge threshold (and then into the
corresponding energy threshold) by means of the threshold scan
procedure. The threshold scan consists in feeding each input channel
of the XAA1.2 with a fixed number of pulses at increasing reference
voltage. With a proper setting of parameters, the counts detected
for each value of the reference voltage will vary from the total
number of pulses (at low threshold) to zero (at high threshold).
Using a linear interpolation, the reference voltage is found such
that the number of detected triggers equals one half of the number
of input pulses. By repeating this procedure at different input
charge values, the reference voltage value as a function of the
input charge can be found, that is the characteristic curve of the
discriminator of each channel of the XAA1.2. This relation can be
used to convert reference voltage values to charge values and to
energy, considering the 3.62 average energy to produce an
electron-hole pair in silicon. Thus the voltage threshold, based on
the background counting rate, can be converted into an energy
threshold using the threshold-energy relation. The SuperAGILE
threshold defined with this method is arbitrary and is mainly useful
to compare conditions where the environmental background is
different.

\begin{figure}[ht]\centering
\includegraphics[width=10 cm, angle=90]{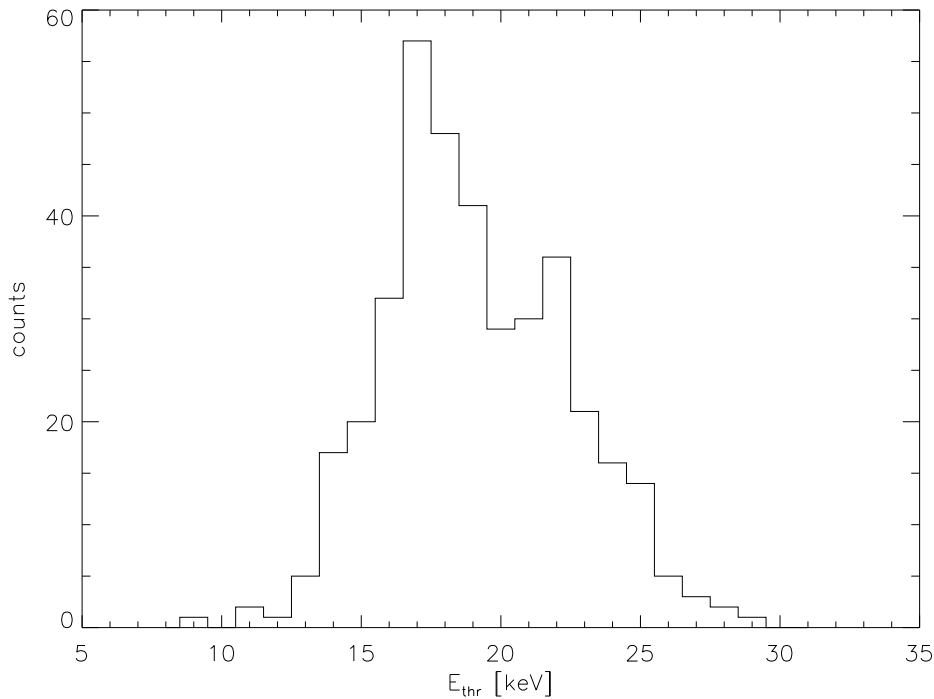}
\caption{Histogram of the energy threshold of the input channels in
a daisy chain.}\label{fig:Ethr_hist}
\end{figure}

As an example, the histogram of the energy threshold of the same
subset of input channels as in sec. \ref{sec:linearity_noise_ecal}
and \ref{sec:linearity_noise_X-rays} and measured using the method
described above is shown in fig. \ref{fig:Ethr_hist}. As it appears
from fig. \ref{fig:Ethr_hist}, the intrinsic threshold of the XAA1.2
channels is largely non uniform, with a typical standard deviation
of 3.3 keV. A fine threshold adjustment is available, based on a
3-bit internal DAC, allowing to reduce the threshold spread.
Unfortunately, the variation of the fine threshold of one channel
affects all the other channels in the same daisy chain, introducing
an offset to their threshold. For this reason, the fine threshold
adjustment is an iterative procedure, starting with the threshold
estimation of each channel with the method described above. The
threshold is then reduced if it is higher and increased if it is
lower than the mean value by changing the digital regulation in the
XAA1.2 configuration mask according to a complex algorithm and the
threshold scan is repeated in the new configuration. Usually in 2--3
steps the threshold spread is minimized.

From preliminary measurements, performed on a sample of 384 channels
(connected to the silicon microstrip detector) with average
threshold value about 20 keV, the fine threshold adjustment reduces
the threshold spread by about a factor of two. A detailed
description of the method and results on the SuperAGILE Flight Unit
will be published elsewhere (\cite{Pacciani_et_al_inprep}).

\section{Optimization of the XAA1.2 performances by means of the
signal shape adjustment} \label{sec:bias_optimization}

The transistor channel noise in the XAA1.2 can be reduced by
increasing the transconductance through an increase of the
\textit{prebias}, controlling the source and the drain current of
the preamplifier. Such a \textit{prebias} variation produces also a
reduction in the peaking time, affecting the shot noise, the thermal
noise in the bias resistors and the noise due to the metal strips,
all depending on the peaking time as reported in sec.
\ref{sec:noise_th}. For SuperAGILE the \textit{prebias} value is
constrained by the chip power consumption, strongly affected by this
bias. In order to keep the peaking time at the 2 $\mu$s nominal
value also the \textit{shabias}, the \textit{Vfp} and the
\textit{Vfs} have to be adjusted.

As an example, by increasing the \textit{prebias} from 50 $\mu$A to
85 $\mu$A and adjusting correspondingly the \textit{shabias},
\textit{Vfp} and \textit{Vfs}, the electronic noise of the XAA1.2
decreases from about 1070 $\mathrm{e^-}$, corresponding to 9.1 keV
FWHM, to about 820 $\mathrm{e^-}$, corresponding to 7.0 keV FWHM
with a power consumption increase of about 20--30 \%.

\begin{figure}[hp!]\centering

\includegraphics[width=10 cm, angle=90]{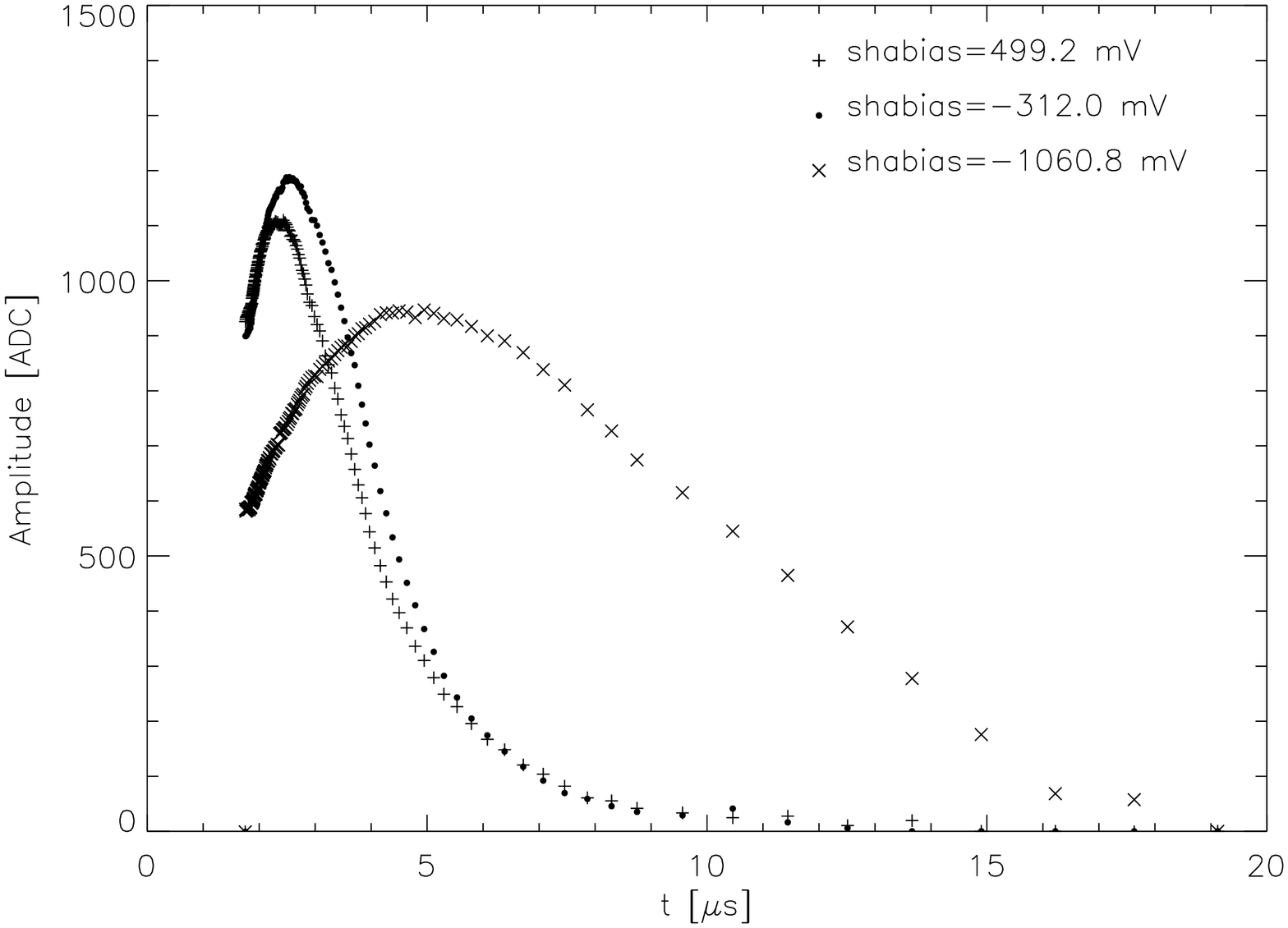}
\caption{Shape of the amplitude signal for three different
\textit{shabias} values. Only the positive amplitudes are shown.}
\label{fig:oscilloscope}

\includegraphics[width=13 cm]{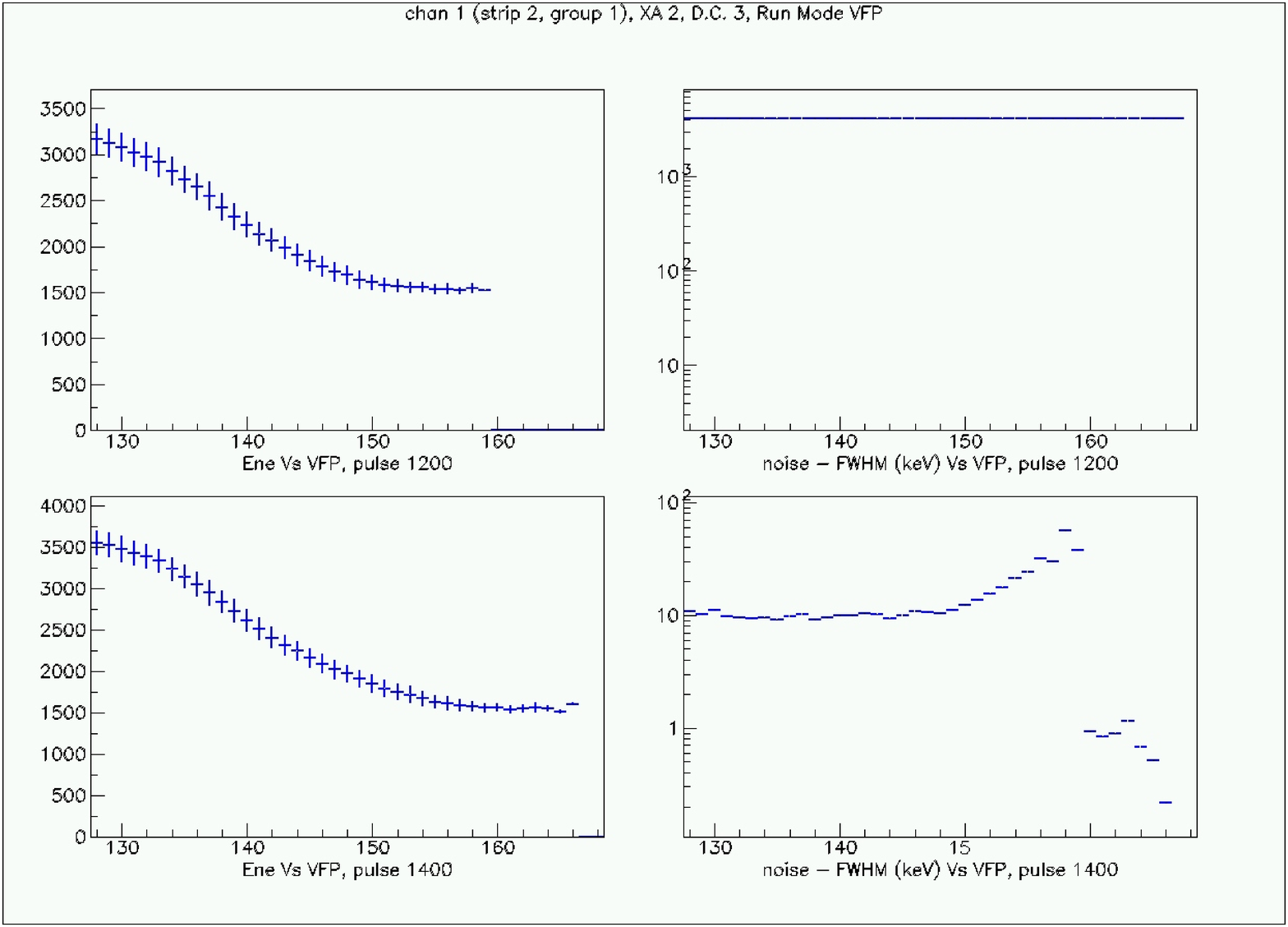}
\caption{Amplitude (left) and noise (right) of a test pulse as a
function of the \textit{Vfp} bias voltage in DAC units (130 digits
corresponding to 31.2 mV and 160 digits to 499.2 mV).}
\label{fig:Vfp_scan}

\end{figure}

The optimization of the signal to noise ratio can be performed as an
iterative procedure involving the two bias currents \textit{prebias}
and \textit{shabias} and the two voltages \textit{Vfp} and
\textit{Vfs}, all affecting the signal shape. Since the XAA1.2 chip
does not allow to access the analogue output signal from the shaper
before the signal is processed by the peak stretcher and S/H, an
indirect way to detect the shaper output signal (in order to control
the effect of the bias currents and voltages variation on the signal
shape) is to switch off the stretcher and to acquire signals at
different \textit{TrigDelbias} values, i.e. with different values of
the delay between the trigger and the sampling time of the internal
S/H, so reconstructing an oscilloscope-like signal shape image, as
shown in fig. \ref{fig:oscilloscope}.

At the beginning of the optimization procedure the \textit{prebias}
is increased. A suitable value for SuperAGILE is about 80--100
$\mu$A, compatible with its power budget. The value of the
\textit{Vfp} is then adjusted to about 100 mV above the value that
produces the preamplifier saturation and the exact bias value is
selected searching the minimum of the noise as a function of
\textit{Vfp} by using the test pulse generator, see for example fig.
\ref{fig:Vfp_scan}. The \textit{shabias} and \textit{Vfs} are then
adjusted to keep the peaking time as close as possible to the 2
$\mu$s nominal value and to reduce the noise. In this way the
minimum of the noise can be found and the optimal values of the
parameters can be selected.

\section{Temperature dependence of XAA1.2 performances}
\label{sec:thermal}

The SuperAGILE instrument in orbit will work between $\mathrm{-20
\degree \; C}$ and $\mathrm{+30 \degree \; C}$ (operative
temperature range). To characterise the temperature dependence of
the XAA1.2 performances, we measured the variation with temperature
of the gain (used to reconstruct the amplitude of the detected
photons) and the address signals (needed to reconstruct the photons
position).

The measurements were performed in a thermo-vacuum chamber, from
$\mathrm{-10 \degree \; C}$ to $\mathrm{+40 \degree \; C}$ at 50
mbar pressure, using the electronic calibration procedure. The SAFEE
PCB is glued to a carbon fiber and aluminum honeycomb tray, the
SuperAGILE collimator and coded mask (carbon fiber and tungsten) are
positioned above the PCB in order to keep the experimental set-up as
close as possible to the true in-orbit conditions. Without the coded
mask and collimator, in vacuum the temperature of the XAA1.2
(measured with an infrared thermometer) is about $\mathrm{5 \div 10
\degree \; C}$ higher than the environment temperature, because of
its power dissipation. The SuperAGILE temperature is monitored using
32 AD590KF thermometers and each SAFEE unit contains two
thermometers, one placed in the horizontal board and the other in
the vertical section. During the test the temperature has been
measured with the thermometer located in the horizontal SAFEE board,
the nearest one to the XAA1.2.

\begin{figure}[hp!]\centering

\includegraphics[width=10 cm, angle=90]{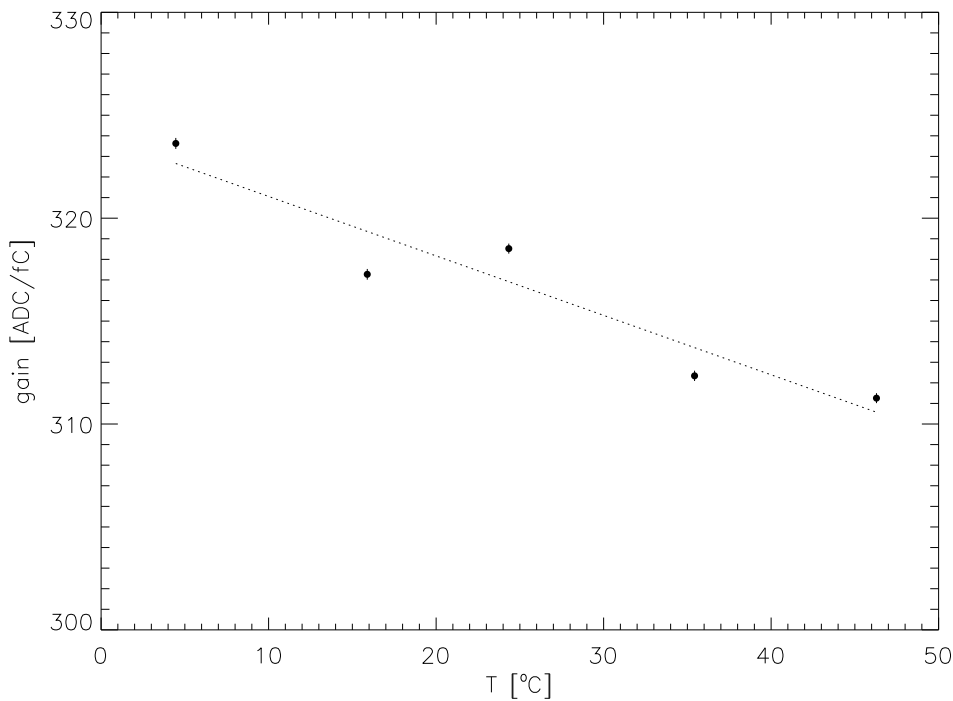}
\caption{Gain (from the calibration curve) as a function of
temperature.} \label{fig:gain_vs_T}

\end{figure}

The gain mean value as a function of temperature between $\mathrm{0
\degree \; C}$ and $\mathrm{40 \degree \; C}$ are shown in fig.
\ref{fig:gain_vs_T}. Below $\mathrm{0 \degree \; C}$ the noise
introduced by the thermo-vacuum chamber strongly affects the
amplitude output signals and does not allow to perform the
electronic calibration procedure. As shown in fig.
\ref{fig:gain_vs_T}, the XAA1.2 gain decreases as a function of
temperature and the percentage variation is $\mathrm{-0.1 \; \%
\cdot (\degree \; C)^{-1}}$ with a linear fit superimposed on the
plot as a dotted line.

\begin{figure}[h]\centering
\includegraphics[width=10 cm, angle=90]{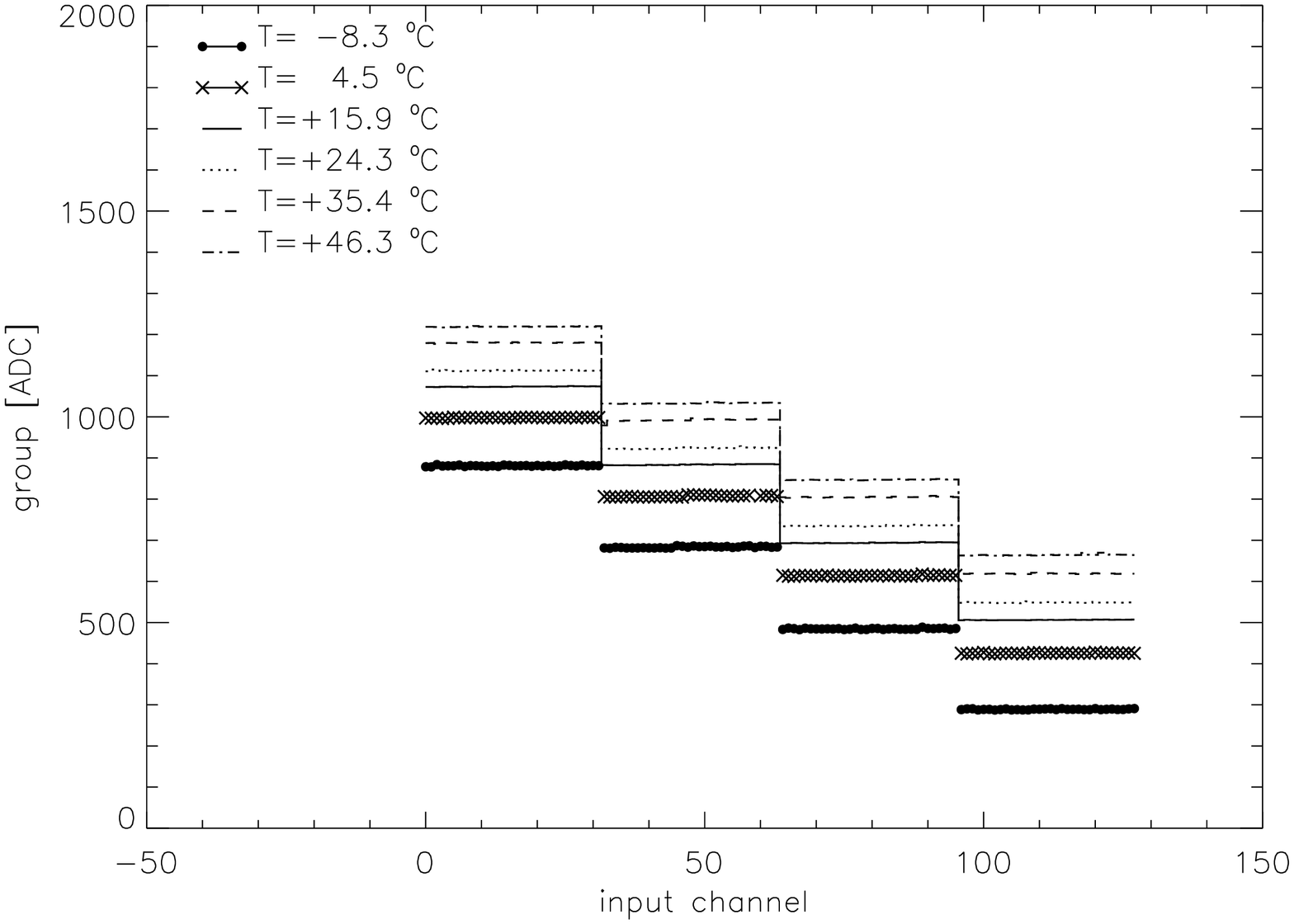}
\caption{Superposition of \textit{group} signals at different
temperatures between $\mathrm{-10 \degree \; C}$ and $\mathrm{+40
\degree \; C}$.}\label{fig:group_vs_T}
\end{figure}

The superposition of the \textit{group} digitized signals in the
range between $\mathrm{-10 \degree \; C}$ and $\mathrm{+40 \degree
\; C}$ is shown in fig. \ref{fig:group_vs_T}. As can be seen in the
plot, the \textit{group} signal shift depends on the temperature,
being about 5 $\mathrm{ADC \; ch.\cdot (\degree C)^{-1}}$ above
$\mathrm{10 \degree \; C}$ and about 10 $\mathrm{ADC \; ch.\cdot
(\degree C)^{-1}}$ below $\mathrm{0 \degree \; C}$. The average
separation between two groups is 200 ADC ch. A temperature variation
of about $\mathrm{20 \degree \; C}$ produces the superposition of
the \textit{group} signal in two consecutive groups of channels (for
example the superposition between the first group at $\mathrm{-10
\degree \; C}$ and the second group at $\mathrm{+10 \degree \; C}$),
that can cause an incorrect reconstruction of the triggering
channel.

\begin{figure}[hp!]\centering

\includegraphics[width=10 cm, angle=90]{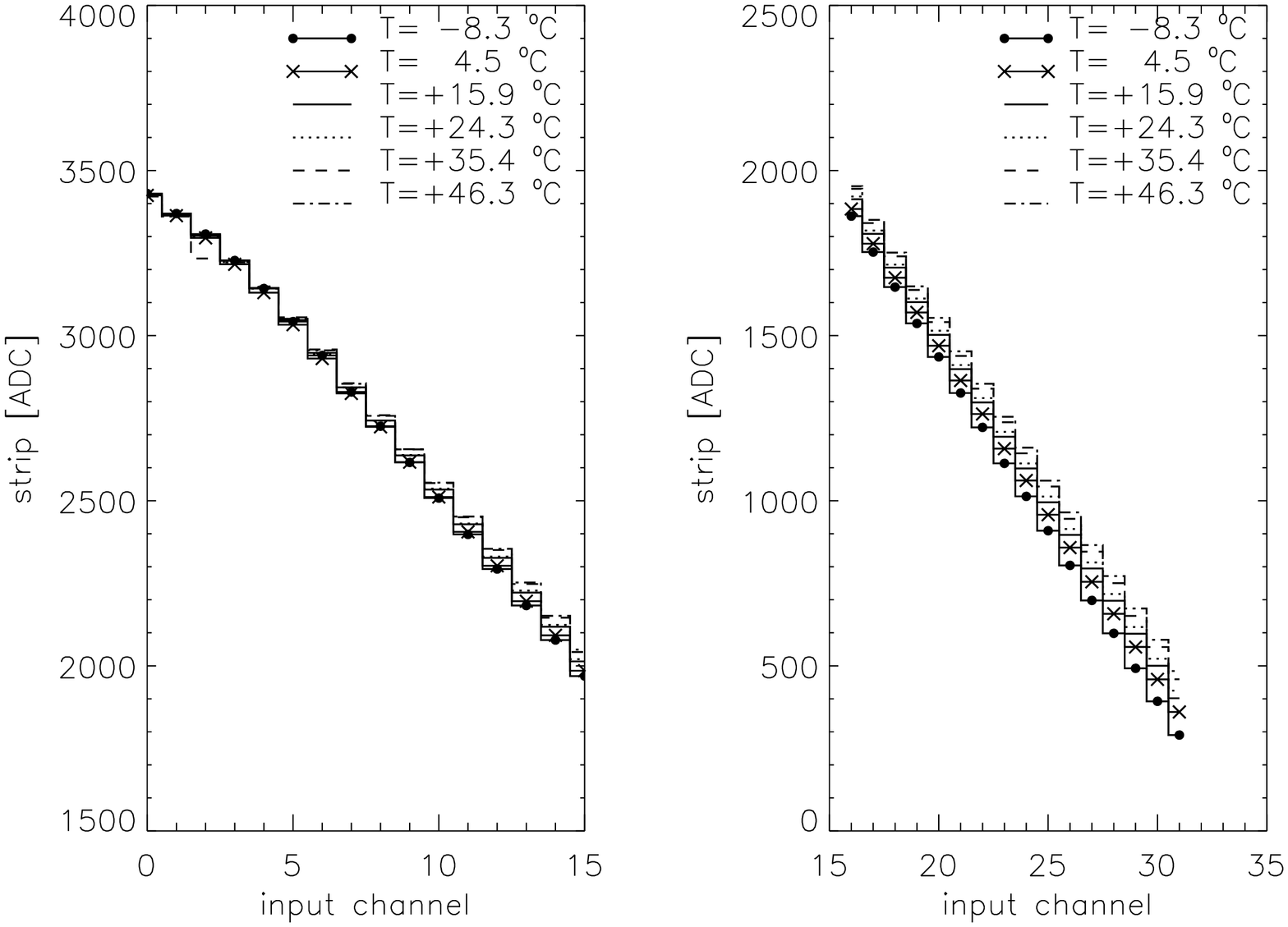}
\caption{Superposition of the \textit{strip} signals in the first
group at different temperatures between $\mathrm{-10 \degree \; C}$
and $\mathrm{+40 \degree \; C}$.}\label{fig:strip_vs_T}

\includegraphics[width=10 cm, angle=90]{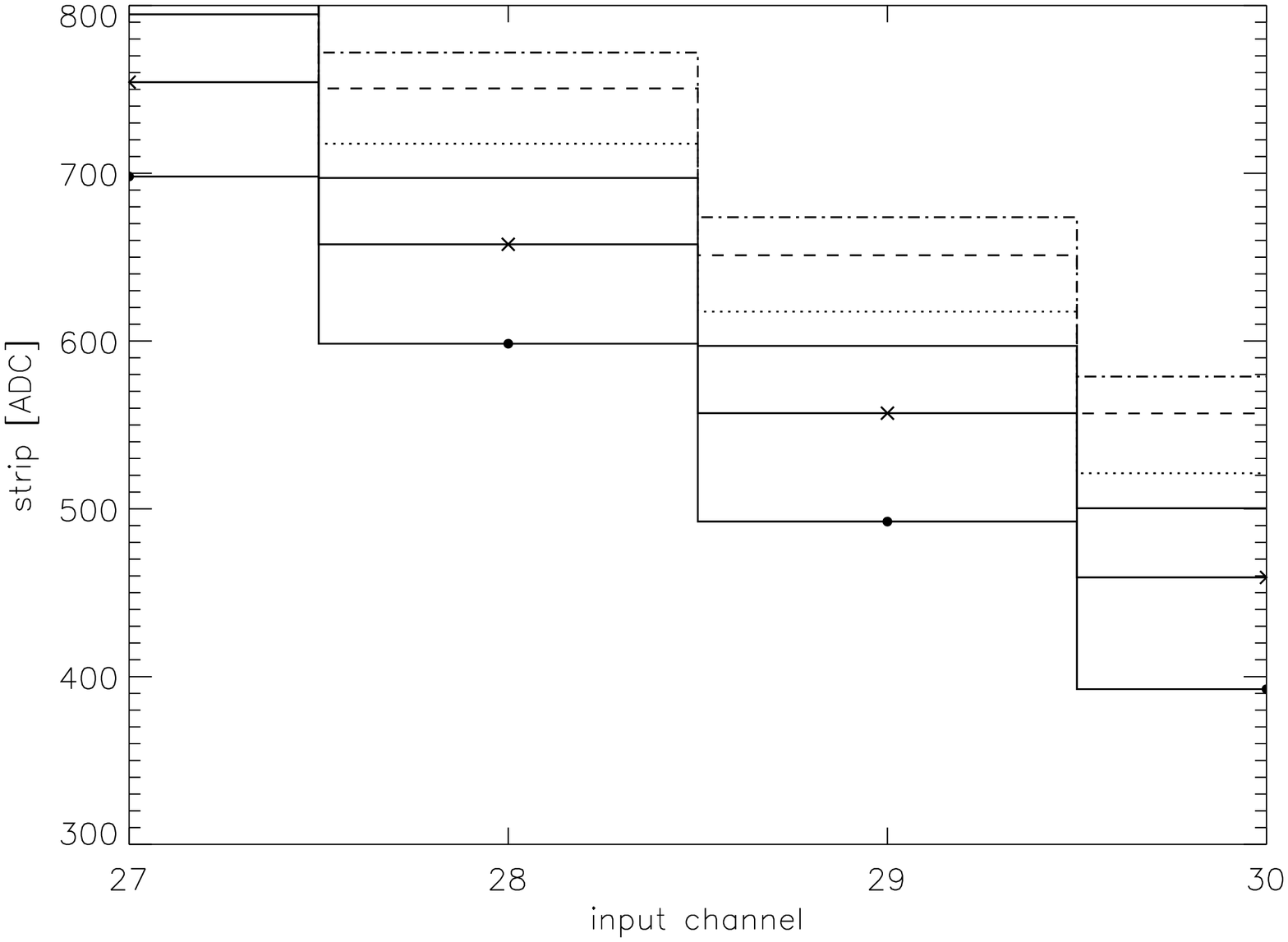}
\caption{Magnified view of the superposition of the strip signals in
the first group (between $\mathrm{-10 \degree \; C}$ and
$\mathrm{+40 \degree \; C}$. The line code is the same as in fig.
\ref{fig:strip_vs_T}.}\label{fig:strip_vs_T_zoom}

\end{figure}

A similar shift as a function of temperature can be seen in the
\textit{strip} signal. A plot of the \textit{strip} digitized signal
of the first group (first 32 channels) is shown in fig.
\ref{fig:strip_vs_T}. For sake of clarity, a magnified view of the
\textit{strip} signal superposition is shown in fig.
\ref{fig:strip_vs_T_zoom}. As can be seen in the plot, the
\textit{strip} variation with temperature depends on the circuit
channel under test, being negligible below channel 15 and important
above channel 25. Particularly, between channel 15 and 25 a shift of
one address channel is seen with a temperature variation of about
$\mathrm{50 \degree \; C}$ while above channel 25 the same shift is
produced by a temperature variation of $\mathrm{20 \degree \; C}$.

\section{Measurement of the XAA1.2 radiation damage}
\label{sec:radiation}

Since the XAA1.2 is not specifically designed as a radiation hard
component for space applications, its radiation hardness required
to be tested. Heavy ion irradiations were performed at the SIRAD
irradiation facility of the Laboratori Nazionali INFN of Legnaro
(see \cite{Wyss_et_al_2001} for a complete description of the
facility) to study the radiation damage of the XAA1.2, mainly
latch-up, SEU and total dose effects.

The XAA1.2 mean LET threshold for the latch-up is $(5.0 \pm 1.5)$
$\mathrm{MeV \cdot cm^2 \cdot mg^{-1}}$ with a mean limit cross
section of $(1.8 \pm 0.1) \cdot 10^{-3}$ cm$^2$ and the LET
threshold for the SEU is $(3.9 \pm 0.3)$ $\mathrm{MeV \cdot cm^2
\cdot mg^{-1}}$. The power consumption of the three XAA1.2 chips
increases above about 100 krad and the amount of the variation
depends on the particular chip considered. From the measurements, at
a typical absorbed dose of 3 krad, variations from $-2 \; \%$ up to
$+2 \; \%$ are estimated. The linearity of the XAA1.2 chip does not
show significant variations at a dose level of 1 krad, some
linearity variations can be seen at a 173 krad level and the chip is
not linear above 700 krad. All the details about the experimental
set-up, data analysis and results were reported by
\cite{Del_Monte_et_al_2005}.

\section{Conclusions}
\label{sec:conclusions}

The XAA1.2 is an ASIC adapted and partially optimized by Ideas ASA
(now Gamma Medica-Ideas) as a front-end electronic chip for 1-D
position sensitive silicon microstrip detectors. The chip is
manufactured with the 0.8 $\mu$m VLSI CMOS double-poly and double
metal technology on epitaxial layer. The circuit is divided into 128
data driven and self-triggering input channels, consisting of an
analogue and a digital part, and is designed to detect single hit
events with sparse readout.

The overall noise in the XAA1.2 chip as a front-end circuit for
silicon microstrip detectors is dominated by the transistor channel
noise. When considering a detector with a 30 $\mathrm{pF \cdot
strip^{-1}}$ capacitance, the estimated noise introduced by the ASIC
is about 935 $\mathrm{e^-}$, corresponding to approximately 8.0 keV
FWHM.

The system gain and noise has been measured using an external test
pulse generator and calibration X-ray sources. Averaged over a daisy
chain of three chips with a total number of 384 input channels, a
noise value of 7.5 keV FWHM is measured with the electronic pulse
generator. Average FWHM values ranging from 7.1 keV (at 22.1 keV) up
to 8.3 keV (at 122.1 keV) are measured using calibration X-ray
sources on the same channels sample and the reported values are
compatible at $4 \sigma$ with the FWHM measured using the test pulse
generator.

The XAA1.2 energy threshold is measured by means of the background
acquisitions and electronic pulse tests. In the same subset of 384
input channels tested with the electronic pulse and the X-ray
sources, the mean threshold is 19.6 keV with a 3.3 keV standard
deviation. The XAA1.2 contains an internal circuit, based on a 3-bit
DAC, to equalize the threshold. From preliminary measurements, the
fine threshold adjustment reduces the threshold spread by about a
factor of two.

The electronic noise of the XAA1.2 chip can be reduced by adjusting
the signal pulse shape considering an optimal peaking time of 2
$\mu$s. The signal shape adjustment involves the bias currents and
voltages of the component and carries an increase in the power
consumption. Using the signal shape optimization method the chip
noise can be reduced down to 7.0 keV FWHM with a power consumption
increase of about 20--30 \%.

The thermal variations affect the linearity and the address signals
of the XAA1.2. Between $\mathrm{0 \degree \; C}$ and $\mathrm{+40
\degree \; C}$ a gain variations of about $\mathrm{-0.1 \; \% \cdot
(\degree \; C)^{-1}}$ is measured. In the temperature range from
$\mathrm{-10 \degree \; C}$ to $\mathrm{+40 \degree \; C}$
significant address variations can be seen on the scale of about
$\mathrm{20 \degree \; C}$.

The XAA1.2 is not designed as a radiation hard component for space
applications. From the radiation damage measurements a mean LET
threshold for the latch-up of $(5.0 \pm 1.5)$ $\mathrm{MeV \cdot
cm^2 \cdot mg^{-1}}$ with a mean limit cross section of $(1.8 \pm
0.1) \cdot 10^{-3}$ cm$^2$ and a LET threshold for the SEU of
$(3.9 \pm 0.3)$ $\mathrm{MeV \cdot cm^2 \cdot mg^{-1}}$ are found.
The power consumption increases with the total ionizing dose above
100 krad while variations on the order of 2 \% are found at a 3
krad dose level. Finally some linearity variations can be seen at
about 200 krad dose while the chip is not linear above 700 krad.

\ack

AGILE is a project funded by ASI, INAF, INFN and CIFS. The authors
want to thank Bj\o rn Sundal (Ideas ASA) for useful suggestion and
discussion about the XAA1.2 ASIC chip features.




\begin{thebibliography}{00}





\bibitem{Dabrowski_2003}
W. {D\c{a}browski}, Nucl. Instr. and Meth. A 501 (2003) 167--174.

\bibitem{Barbiellini_1987}
G. {Barbiellini}, Proceedings of Physics and Astrophysics in the
Space Station Era, P. L. Bernacca, R. Ruffini (Eds.), Venice 4-7
October 1987.

\bibitem{Barbier_et_al_1998}
L. M. {Barbier}, E. {Karageorge}, S. {Singh}, Nucl. Instr. and Meth.
A 417 (1998) 354--359.

\bibitem{do_Couto_2001}
E. {do Couto e Silva}, Nucl. Instr. and Meth. A 473 (2001) 107--113.

\bibitem{Boezio_et_al_2004}
M. {Boezio}, V. {Bonvicini}, E. {Mocchiutti}, P.  {Schiavon}, A.
{Vacchi}, G. {Zampa}, N. {Zampa}, A. {Bakaldin}, A.~M. {Galper},
S.~V. {Koldashov}, M.~G. {Korotkov}, V.~V. {Mikhailov}, S.~A.
{Voronov}, Y. {Yurkin}, A. {Basili}, R. {Bencardino}, L.
{Bongiorno}, M. {Casolino}, M.~P. {de Pascale}, G. {Furano}, A.
{Menicucci}, M. {Minori}, A. {Morselli}, P. {Picozza}, R.
{Sparvoli}, R. {Wischnewski}, O. {Adriani}, L. {Bonechi}, M.
{Bongi}, F. {Giambi}, P. {Papini}, S.~B. {Ricciarini}, P.
{Spillantini}, S. {Straulino}, F. {Taccetti}, E. {Vannuccini}, G.
{Castellini}, P. {Carlson}, J. {Lund}, J. {Lundquist}, S. {Orsi}, M.
{Pearce}, G.~C. {Barbarino}, D. {Campana}, G. {Osteria}, G. {Rossi},
S. {Russo}, M. {Boscherini}, W. {Menn}, M. {Simon}, M. {Ricci}, M.
{Ambriola}, R. {Bellotti}, F. {Cafagna}, M. {Circella}, C. {de
Marzo}, N. {Giglietto}, M. {Mirizzi}, M. {Romita}, P. {Spinelli}, E.
{Bogomolov}, S. {Krutkov}, G. {Vasiljev}, G. {Bazilevskaja}, A.
{Grigorjeva}, R. {Mukhametshin}, Y. {Stozhkov}, J.~W. {Mitchell},
R.~E. {Streitmatter}, S.~J. {Stochaj}, Nucl. Phys. B  134 (2004)
{39-46}.

\bibitem{Tavani_et_al_2006}
M. {Tavani}, G. {Barbiellini}, A. {Argan}, M. {Basset}, F.
{Boffelli}, A. {Bulgarelli}, P. {Caraveo}, A. {Chen}, E. {Costa}, G.
{De Paris}, E. {Del Monte}, G. {Di Cocco}, I. {Donnarumma}, M.
{Feroci}, M. {Fiorini}, L. {Foggetta}, T. {Froysland}, M. {Frutti},
F. {Fuschino}, M. {Galli}, F. {Gianotti}, A. {Giuliani}, C.
{Labanti}, I. {Lapshov}, F. {Lazzarotto}, F. {Liello}, P. {Lipari},
F. {Longo}, M. {Marisaldi}, M. {Mastropietro}, E. {Mattaini}, F.
{Mauri}, S. {Mereghetti}, E. {Morelli}, A. {Morselli}, L.
{Pacciani}, A. {Pellizzoni}, F. {Perotti}, P. {Picozza}, C.
{Pittori}, C. {Pontoni}, G. {Porrovecchio}, M. {Prest}, G.
{Pucella}, M. {Rapisarda}, E. {Rossi}, A. {Rubini}, P. {Soffitta},
A. {Traci}, M. {Trifoglio}, A. {Trois}, E. {Vallazza}, S.
{Vercellone}, A. {Zambra}, D. {Zanello}, Proc. SPIE 6266 (2006)
626603-1.

\bibitem{Ritz_et_al_2004}
S. {Ritz}, P.~F. {Michelson}, C. {Meegan}, J. {Grindlay}, GLAST
Mission, Bulletin of the American Astronomical Society 205 (2004)
606.

\bibitem{Prest_et_al_2003}
M. {Prest}, G. {Barbiellini}, G. {Bordignon}, G. {Fedel}, F.
{Liello}, F. {Longo}, C. {Pontoni}, E. {Vallazza}, Nucl. Instr. and
Meth. A 501~(1) (2003) 280--287.

\bibitem{Labanti_et_al_2006}
C. {Labanti}, M. {Marisaldi}, F. {Fuschino}, M. {Galli}, A. {Argan},
A. {Bulgarelli}, E. {Costa}, G. {Di Cocco}, F. {Gianotti}, M.
{Tavani}, M. {Trifoglio}, Proc. SPIE 6266 (2006) 62663Q-1.

\bibitem{Soffitta_et_al_2006}
P. {Soffitta}, E. {Costa}, E. {Del Monte}, G. {Di Persio}, I.
{Donnarumma}, Y. {Evangelista}, M. {Feroci}, M. {Frutti}, I.
{Lapshov}, F. {Lazzarotto}, M. {Mastropietro}, E. {Morelli}, L.
{Pacciani}, G. {Porrovecchio}, M. {Rapisarda}, A. {Rubini}, G.
{Sabatino}, O. {Uberti}, A. {Argan}, M. {Tavani}, Proc. SPIE 6266
(2006) 626631-1.

\bibitem{Perotti_et_al_2006}
F. {Perotti}, M. {Fiorini}, S. {Incorvaia}, E. {Mattaini}, E.
{Sant'Ambrogio}, Nucl. Instr. and Meth. A 556 (2006) 228--236.

\bibitem{Feroci_et_al_inprep} M. Feroci et al., in preparation.

\bibitem{Fenimore_Cannon_1978}
E.~E. {Fenimore}, T.~M. {Cannon}, Appl. Opt. 17 (1978) 337--347.

\bibitem{Zand_1992}
J. J.~M. {in't Zand}, Ph.D. thesis, University of Utrecht (1992)
ISBN 90-393-0473-4.

\bibitem{Soffitta_et_al_2000}
P. {Soffitta}, E. {Costa}, E. {Del Monte}, M. {Feroci}, I.
{Lapshov}, M. {Mastropietro}, E. {Morelli}, M. {Rapisarda}, A.
{Rubini}, G. {Barbiellini}, F. {Longo}, S. {Mereghetti}, A.
{Morselli}, M. {Prest}, M. {Tavani}, E. {Vallazza}, S. {Vercellone},
Proc. SPIE 4140 (2000) 283--292.

\bibitem{Del_Monte_et_al_2000}
E. {Del Monte}, E. {Costa}, G. {Di Persio}, M. {Feroci}, I.
{Lapshov}, B. {Martino}, M. {Mastropietro}, E. {Morelli}, M.
{Prest}, A. {Rubini}, P. {Soffitta}, E. {Vallazza}, Proc. SPIE 4140
(2000) 584--594.

\bibitem{Pacciani_et_al_2006}
L. {Pacciani}, E. {Morelli}, A. {Rubini}, M. {Mastropietro}, G.
{Porrovecchio}, E. {Costa}, E. {Del Monte}, I. {Donnarumma}, Y.
{Evangelista}, M. {Feroci}, F. {Lazzarotto}, M. {Rapisarda}, P.
{Soffitta}, submitted to Nucl. Instr. and Meth. A (2006).

\bibitem{Barichello_et_al_1998}
G. {Barichello}, A. {Cervera-Villanueva}, D.~C. {Daniels}, E. {do
Couto e Silva}, L. {Dumps}, M.~{Ellis}, D. {Ferr\`{e}re}, J.~J.
{Gomez-Cadenas}, M. {Gouan\`{e}re}, J.~A. {Hernando}, W. {Huta},
J.~M. {Jimenez}, J. {Kokkonen}, V.~E. {Kuznetsov}, L. {Linssen}, B.
{Lisowski}, {\" O}. {Runolfsson}, F. J.~P. {Soler}, D. {Steele}, M.
{Stip\v{c}evij\'{c}}, M. {Veltri}, Nucl. Instr. and Meth. A. 413
(1998) 17--30.

\bibitem{Knoll_1989}
G.~F. {Knoll}, {Radiation detection and measurement}, 2nd Edition,
John Wiley and Sons, (New York, 1989).

\bibitem{Zombeck}
M.~V. Zombeck, {Handbook of Astronomy and Astrophysics}, 2nd
Edition, Cambridge University Press (Cambridge, UK, 1990).

\bibitem{Pacciani_et_al_inprep} L. Pacciani et al., in preparation.

\bibitem{Wyss_et_al_2001}
J. {Wyss}, D. {Bisello}, D. {Pantano}, Nucl. Instr. and Meth. A 462
(2001) 426--434.

\bibitem{Del_Monte_et_al_2005}
E. {Del Monte}, L. {Pacciani}, G. {Porrovecchio}, P. {Soffitta}, E.
{Costa}, G. {Di Persio}, M. {Feroci}, M. {Mastropietro}, E.
{Morelli}, M. {Rapisarda}, A. {Rubini}, D. {Bisello}, A.
{Candelori}, A. {Kaminski}, J. {Wyss}, Nucl. Instr. and Meth. A 538
(2005) 465--482.

\end{thebibliography}
\end{document}